\documentclass[10pt,aps,prc,floatfix,twocolumn,nofootinbib, superscriptaddress,preprintnumbers]{revtex4-1}
\usepackage{amsfonts,amsmath,amssymb,bm, xspace}
\usepackage{color,graphicx}
\usepackage{bbm}
\usepackage{dcolumn}                 
\usepackage{graphicx,amsmath}
\usepackage{color}
\usepackage{amssymb}
\usepackage{hyperref} 
\usepackage{float}
\usepackage[bottom]{footmisc}

\usepackage{tikz}
\usetikzlibrary{shapes}
\usepackage{pgfplots}
\usepackage{wrapfig}

\usepackage{natbib}
\usepackage{graphicx}

\newcommand{\orcid}[1]{\href{https://orcid.org/#1}{\includegraphics[width=20pt]{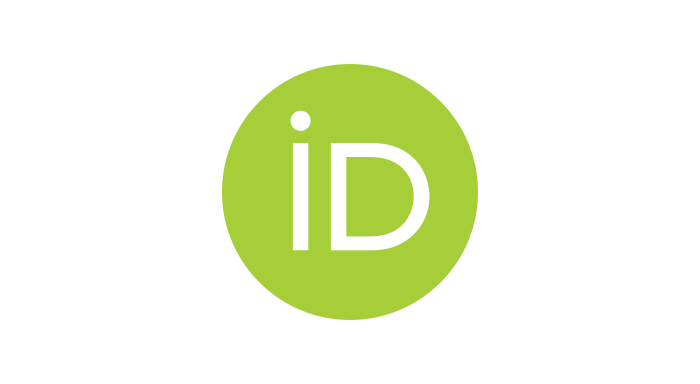}}}
\newcommand{\sm}{\mathrm{SM}}
\newcommand{\nm}{\mathrm{NM}}
\newcommand{\nq}{\mathrm{NQ}}

\newcommand{\FS}{\mathrm{FS}}
\newcommand{\um}{\mathrm{UM}}

\newcommand{\cldm}{\mathrm{CLDM}}
\newcommand{\ecldm}{\mathrm{eCLDM}}
\newcommand{\dd}{\mathrm{DD}}

\newcommand{\cl}{\mathrm{cl}}
\newcommand{\eq}{\mathrm{eq}}

\newcommand{\xt}{\tilde{x}}

\newcommand{\sat}{\mathrm{sat}}
\newcommand{\sym}{\mathrm{sym}}

\newcommand{\coul}{\mathrm{Coul}}
\newcommand{\surf}{\mathrm{surf}}

\newcommand{\gmr}{\mathrm{ISGMR}}

\newcommand{\bat}{\Bigr\rvert}

\begin{document}

\title{Nuclear incompressibility and sound speed in uniform matter and finite nuclei}

\author{G. Grams\orcid{0000-0002-8635-383X}}
\email{guilherme.grams@ulb.be}
\affiliation{Univ Lyon, Univ Claude Bernard Lyon 1, CNRS/IN2P3, IP2I Lyon, UMR 5822, F-69622, Villeurbanne, France}
\affiliation{Institut d’Astronomie et d’Astrophysique, CP-226, Université Libre de Bruxelles, 1050 Brussels, Belgium}

\author{R. Somasundaram\orcid{0000-0003-0427-3893}}
\email{r.somasundaram@ip2i.in2p3.fr}
\affiliation{Univ Lyon, Univ Claude Bernard Lyon 1, CNRS/IN2P3, IP2I Lyon, UMR 5822, F-69622, Villeurbanne, France}
 
\author{J. Margueron\orcid{0000-0001-8743-3092}}
\affiliation{Univ Lyon, Univ Claude Bernard Lyon 1, CNRS/IN2P3, IP2I Lyon, UMR 5822, F-69622, Villeurbanne, France}

\author{E. Khan\orcid{0000-0002-1343-7805}}
\affiliation{IJCLab, Universit\'e Paris-Saclay, CNRS/IN2P3, 91405 Orsay Cedex, France}

\date{\today}

\begin{abstract}
We have extended the compressible liquid-drop model (CLDM) with a density-dependent surface term (eCLDM), which allows
for a unified description of both the nuclear ground state energies and the incompressibility modulus in finite nuclei $K_A$. We analyse the role of the nuclear empirical parameters, e.g., $K_\sat$, $Q_\sat$, $L_\sym$ and $K_\sym$, which contribute to the bulk properties, as well as the role of the finite size contributions. For the bulk properties, the density and isospin dependencies of the nuclear incompressibility in infinite matter are  characterized by introducing new empirical parameters, and two new constraints for the value of $K_\sym$ are suggested. For finite nuclei, we employ a Bayesian approach coupled to a Markov-Chain Monte-Carlo (MCMC) exploration of the parameter space to confront the model predictions of $K_A$ in Zr, Sn and Pb isotopes to the experimental data. We show that $Q_\sat\approx -950\pm200$~MeV describes the experimental measurements of $K_A$ in these isotopes. This value is different from the ones deduced from phenomenological nuclear energy density functionals, suggesting a possible explanation of their difficulty to accurately describe Zr, Sn and Pb data all together. In addition we explore the impact of a fictitious measurement of the Giant Monopole Resonance energy in $^{132}$Sn. We show that this measurement, provided it is accurate enough, will allow to better determine $K_\sym$ and $K_\tau$. Finally we explore the properties of the sound speed around saturation density and show the important role of finite size terms in finite nuclei since they reduce the sound speed to approximately half compared to nuclear matter. 
\end{abstract}

\maketitle
\section{Introduction}

The response of nuclear matter to compression and expansion plays a very important role in many phenomena in nature, from finite nuclei~\cite{HarakehWoude:2001}, which can be viewed as non-uniform pieces of nuclear matter squeezed by the effects of the surface terms, up to astrophysical nuclear systems such as neutron stars, supernovae or kilonovae~\cite{Rezzolla2018}, where nuclear matter explores densities and isospin asymmetries in extreme regimes. In finite nuclei, the repulsive surface tension and the Coulomb interaction counter balance the attractive bulk nuclear force and allow the exploration of densities close to the saturation density of nuclear matter ($n_\sat\approx 0.155$~fm$^{-3}$~\cite{meta1}), while in compact stars, the bulk nuclear force resists gravity for densities corresponding to several times saturation density. In these examples, the equilibrium states of these systems represent a balance between the bulk properties and the action of external forces (finite size terms or gravitational force). It is then important to quantify precisely the response of bulk nuclear matter (incompressibility) from analyses of finite nuclei properties (giant monopole resonances), which is the scope of the present study.

At first order, the energy required to compress matter from its equilibrium state, is given by the incompressibility modulus $K_\sat$ with an isospin asymmetry dependence driven by the parameter $K_\tau$~\cite{Blaizot80}: $K_\sat+\delta^2 K_\tau$. These nuclear empirical parameters could be extracted from the analysis of the isoscalar giant monopole resonance (ISGMR), excited by the scattering of alpha particles, see for instance Ref.~\cite{GARG2018} and references therein. The relation between the energy of the ISGMR, $E_\gmr$, and the incompressibility in finite nuclei $K_A$ is~\cite{GARG2018}, 
\begin{equation}
E_\gmr = \hbar \sqrt{\frac{K_A}{m_N\langle r^2\rangle}} \simeq \hbar \sqrt{\frac{5 K_A}{3m_N R_A}}\, ,
\label{eq:egmr}
\end{equation}
where $m_N$ is the nucleon mass and $\langle r^2\rangle$ is the mean square radius of the density distribution in finite nuclei. The last expression is obtained assuming a flat density distribution up to $R_A$, as in the compressible liquid-drop model (CLDM). Considering a leptodermous expansion as in the liquid-drop model, the incompressibility modulus in finite nuclei $K_A$ can be expressed as~\cite{Blaizot80}
\begin{equation}
K_A = K_\sat + K_\tau\delta^2 + K_\coul \frac{Z^2}{A^{4/3}} + K_\surf A^{-1/3} + \dots
\label{eq:kaemp}
\end{equation}
where $K_\sat$ and $K_\tau$ are the bulk contributions which we aim to extract from experimental data, $K_\coul$ is the Coulomb repulsive contribution and $K_\surf$ the surface attractive contribution. This leptodermous expansion is however difficult to employ for the determination of $K_\sat$ and $K_\tau$ from experimental measurements of $E_\gmr$, as Eqs.~\eqref{eq:egmr} and \eqref{eq:kaemp} may suggest, especially since the term $K_\surf$ is difficult to fix from the few existing experimental data. The situation is different in the case of the leptodermous expansion of the nuclear mass, since more than 2000 nuclei have been measured~\cite{AME2020}. In a recent analysis~\cite{BaoAnLie21}, precise values for $K_\sat$ and $K_\tau$ have been obtained from Eq.~\eqref{eq:kaemp}, fixing $K_\surf=cK_\sat$ (with $c\approx -1.2 \pm 0.12$ \cite{Patra2002}) and $K_\coul \approx -5.2 \pm 0.7$ MeV \cite{Sagawa2007}. This indicates that the key quantity which would allow the use of such an empirical relation is the surface term. In this paper, we investigate the impact of the surface term in the CLDM framework and its role to reproduce experimental data.

By using the energy density functional approach, the first precise extraction of $K_\sat$ gave $K_\sat=210\pm30$~MeV~\cite{Blaizot80}, corrected to $240\pm20$~MeV later on as a good compromise between data in $^{208}$Pb and $^{90}$Zr~\cite{GARG2018}. The isospin dependence of the incompressibility $K_\tau$ is more difficult to determine from experimental data. It has recently been extracted from systematical exploration of Sn isotopic chain, giving $K_\tau\approx -550\pm100$~MeV~\cite{GARG2018}. However, there are several unsettled questions: by using non-magic nuclei, the analysis of the data requires the understanding of many-body correlations (pairing, deformation, etc.) on the incompressibility of finite nuclei. The question of the isoscalar and isovector properties of the incompressibility is also important since the density and the isospin asymmetry distributions in finite nuclei are different from one to another. In addition, a systematic difference between the incompressibility extracted from $^{208}$Pb and from $^{120}$Sn (which tends toward $K_\sat\simeq 205$ MeV\cite{avo13}) remains which origin is still not well understood. This issue could be of similar origin as the systematical dispersion for $K_\sat$ obtained by using different models, where $K_\sat\approx 220$~MeV is preferred by Gogny forces \cite{gor09} while non-linear relativistic mean field models favor $K_\sat\approx 250$~MeV \cite{Lalazissis05}. It was suggested that these systematical differences could be related to the different density dependence of the models, encoded in the nuclear empirical parameter (NEP) $Q_\sat$~\cite{Khan2012,Khan2013b}. It is indeed a general result that a large uncertainty on a high order NEP impacts the precise determination of lower order ones~\cite{Margueron2019}. 

In uniform matter (UM), the incompressibility $K_\um$ is defined as the second derivative of the energy density $\epsilon_\um=E_\um/V$ as
\begin{eqnarray}
K_\um(n,\delta) &=& 9 n \frac{\partial^2 \epsilon_\um(n,\delta)}{\partial n^2} \, ,\label{eq:kev} \\
&=&\frac{18}{n} P_\um(n,\delta) + 9 n^2 \frac{\partial^2 e_\um(n,\delta)}{\partial n^2} \, ,
\label{eq:kea}
\end{eqnarray}
where $n$ is the isoscalar density $n=n_n+n_p$ and $\delta$ the isovector parameter $\delta=(n_n-n_p)/n$, the energy per particle is $e_\um=\epsilon_\um/n$, and the pressure $P_\um$ is defined as
\begin{equation}
P_\um = n^2 \frac{\partial e_\um}{\partial n} \, .
\label{eq:pum}
\end{equation}
Note that $K_\um=K_\sat$ if $n=n_\sat$ and $\delta=0$. In finite nuclei, the isovector parameter is noted $\delta_A=(N-Z)/A$.

In the absence of external forces, such as gravity for instance, matter minimizes its energy (mechanical equilibrium) by imposing $P_\um=0$. We note $n_\eq^\um$ the equilibrium density in symmetric (SM) and isospin asymmetric (AM) matter. The latter always deals with small isospin asymmetries $\vert\delta_A\vert\lesssim 0.3$ as expected in finite nuclei. In nuclear matter and at equilibrium, the first term in Eq.~\eqref{eq:kea} vanishes but in finite nuclei however the equilibrium density $n_\eq^A$ is slightly different from the one in uniform matter $n_\eq^\um$, due to the presence of finite size terms which contribute to the pressure. This effect shifts $n_\eq^\um$ by about 10\% at maximum and impacts the value of the bulk incompressibility in finite nuclei. One could then view the finite size terms as an "external" force probing the response of the bulk. Consequently, there is a contribution of the finite size terms to the incompressibility in finite nuclei, in addition to the density and isospin asymmetry dependence of the bulk term~\cite{Blaizot80}. In addition, the equilibrium density $n_\eq^A$ in finite nuclei varies around $n_\eq^\um$ through the nuclear chart, modifying the value of the energy in the bulk. Since this value is controlled at first order by the incompressibility modulus, the energy of finite nuclei in their ground state also contains a contribution originating from the incompressibility of nuclear matter, in addition to the symmetry energy and to the finite size terms. This contribution is difficult to extract from microscopic approaches, e.g., energy density functional, shell model approaches, as well as ab-initio ones, but it could be more visible in macroscopic models such the CLDM that we employ in this study. The fact that the fluctuations in $n_\eq^A$ impact both the energy $e_A$ and the incompressibility $K_A$ requires to employ a model which could describe these two quantities in a unique framework. This is the motivation for the development of the eCLDM that we present in this paper.

The CLDM has been shown relevant to describe nuclear masses~\cite{Myers69,Weiss69} and was also employed to study the clusterized matter present on neutron star crusts \cite{bbp1971,Steiner2008,Carreau2019a,Grams2022a,Grams2022b}. Many variations of the model can be found in the literature, however, it has been argued by Blaizot~\cite{Blaizot80} that the CLDM is not appropriate to accurately extract the incompressibility modulus $K_\sat$ from finite nuclei.  The reason lies in the contribution of the density dependent surface term to the incompressibility, which is absent in most of the macroscopic models. In the present work, we however construct an extended CLDM (eCLDM) with a density-dependent surface tension allowing to describe both nuclear masses and incompressibilities. Furthermore, the bulk term of the present model is described with the meta-model \cite{meta1}, an energy density functional in which the parameters of the model are the empirical parameters of nuclear matter. The meta-model has the advantage of being flexible enough to allow an independent variation of the NEP and can thus be used to easily perform a sensitivity analysis of the individual impact of the NEP on the incompressibility $K_A$, as well as extensive searches of the best parameter sets reproducing experimental data.

The paper is organized as follow: In Sec.~II we explore the incompressibility modulus in nuclear matter in terms of the NEP, or equivalently as a function of the density and the isospin asymmetry. A new constraint on $K_\sym$ is derived and compared to other existing ones. Following the line suggested by Blaizot~\cite{Blaizot80}, we then address finite nuclei in Sec.~III as described by our eCLDM model (with a density dependent surface tension), which allows to reproduce both finite nuclei and incompressibility modulus from the same approach. In Sect.~IV we compare the predictions of the eCLDM to experimental data and analyse the role of the NEP $K_\sat$, $Q_\sat$, $L_\sym$ and $K_\sym$ in a Bayesian framework. Finally, in Sec.~V we discuss the sound speed in both uniform matter and finite nuclei.

\section{Uniform matter}

In this section, we briefly summarize the present understanding of uniform matter and show how the knowledge of the NEP could be used to explore its properties around saturation density. We also present an alternative representation where the reference density is taken to be $n_\eq^{UM}$, the equilibrium density which is a function of $\delta$, instead of the saturation density $n_\sat$ in the usual approach.

\subsection{Representation of the nuclear matter properties in terms of the nuclear empirical parameters}

The NEP, e.g., $E_\sat$, $E_\sym$, are defined as the coefficients of the series expansion of the energy per particle in SM ($e_\sm$) and of the symmetry energy ($e_\sym$) as,
\begin{eqnarray}
e_{\sm} (n) &=& E_\sat + \frac 1 2 K_\sat x^2 + \frac{1}{6} Q_\sat x^3  \nonumber \\ &&\hspace{1cm}+ \frac{1}{24} Z_\sat x^4+\dots \, , \label{eq:eSMEP} \\
e_{\sym}(n) &=& E_\sym + L_\sym x  + \frac 1 2 K_\sym x^2 + \frac{1}{6} Q_\sym x^3 \nonumber \\ 
&&\hspace{1cm}+ \frac{1}{24} Z_\sym x^4+\dots \, ,
\label{eq:eSymEP}
\end{eqnarray}
where $x=(n-n_\sat)/3n_\sat$, with $n_\sat$ being the saturation density of nuclear matter ($n_\sat= 0.155\pm0.005$~fm$^{-3}$, see for instance Ref.~\cite{meta1}). Note that choosing $n_\sat$ as the reference density for the parameter $x$, is arbitrary: in Sec.~\ref{sec:ealt} for instance, we explore another reference density. It should also be noted that in Eq.~\eqref{eq:eSymEP}, the symmetry energy is defined as the difference between neutron matter (NM) and SM energies, as $e_{\sym}(n)=e_{\nm}(n) -e_{\sm}(n)$. It can be expanded in terms of $\delta^2$ as $e_{\sym}(n)=e_{\sym,2}(n)\delta^2 + e_\nq$, where $e_{\sym,2}$ and $e_\nq$ subsume the quadratic and non-quadratic (NQ) contributions respectively.

It was suggested in Ref.~\cite{meta1} to consider the series expansion up to order 4 in the density parameter $x$ in order to represent accurately the energy per particle, the pressure and the sound speed of existing models up to about $4n_\sat$. We adopt this prescription here as well, even if we do not explore such high densities.

Note that since asymmetric matter is mostly quadratic in $\delta$, as it is expected to be~\cite{Somasundaram2021}, Eqs.~\eqref{eq:eSMEP}-\eqref{eq:eSymEP} could also be written in a more compact way,
\begin{eqnarray}
e_\um(x,\delta) &\approx& e_{\sm} (n) + e_{\sym}(n)\delta^2 \, , \\
&\approx& E(\delta) + L_\sym x\delta^2 + \frac 1 2 K(\delta)x^2 + \frac 1 6 Q(\delta)x^3\nonumber \\
&&\hspace{2cm}+\frac 1 {24} Z(\delta)x^4+\dots\, ,
\label{eq:eAMEP}
\end{eqnarray}
where 
\begin{eqnarray}
E(\delta)&\equiv&E_\sat+E_\sym\delta^2 \, ,
\hspace{0.3cm}K(\delta)\equiv K_\sat+K_\sym\delta^2 \, ,\\
Q(\delta)&\equiv&Q_\sat+Q_\sym\delta^2 \, ,
\hspace{0.3cm}Z(\delta)\equiv Z_\sat+Z_\sym\delta^2 \, .
\end{eqnarray}

It should be noted that the above expression of K($\delta)$ is by no means the true isospin dependence of the incompressibility, as it will be discussed below. In particular, it neglects the contribution of the pressure which is different from zero as one gets farther from saturation. It solely represents the second order term in the density expansion of the energy per particle. 

From Eq.~\eqref{eq:eAMEP}, one could deduce a similar expression for the energy density $\epsilon_\um= (1+3x) e_\um n_\sat$ as,
\begin{eqnarray}
\epsilon_\um(x,\delta)/n_\sat &=& E(\delta) + 
L^\epsilon(\delta) x + \frac 1 2 K^\epsilon(\delta)x^2 + \dots\, ,
\label{eq:eVMEP}
\end{eqnarray}
where
\begin{eqnarray}
L^\epsilon(\delta) &\equiv& 3E_\sat+(3E_\sym+L_\sym)\delta^2 \, , \\
K^\epsilon(\delta) &\equiv& K_\sat + K_\sym^\epsilon\delta^2 \, ,
\end{eqnarray}
with
\begin{eqnarray}
K^\epsilon_\sym &\equiv& K_\sym + 6 L_\sym \, .
\label{eq:ksymprime}
\end{eqnarray}
The $\delta$-dependence of the energy density curvature $K^\epsilon(\delta)$ is different from that of the energy per particle curvature $K(\delta)$. Consequences will be discussed in the following, especially for the incompressibility modulus in asymmetric matter. It will be shown that $K^\epsilon(\delta)$ do correspond to the isospin dependence of the incompressibility around saturation density, contrarily to K$_\sym$, which is only a parameter useful in the expansion (\ref{eq:eAMEP})

\begin{table*}[t]
\centering
\tabcolsep=0.4cm
\def\arraystretch{1.5}
\begin{tabular}{cccccccccc}
\hline\hline
Model                & BSK14 & BSK16 & F0 & LNS5 & RATP & SGII & SKI2 & SKO & SLy5 \\
Ref. & \cite{Goriely07} & \cite{Chamel08} & \cite{Lesinski06} & \cite{Cao06} & \cite{Rayet82} & \cite{Nguyen81} & \cite{REINHARD95} & \cite{Reinhard99} & \cite{Chabanat98} \\
\hline
$E_\sat$ (MeV)      &  -15.85 & -16.05 & -16.03 & -15.56  &  -16.05  & -15.59 & -15.76  & -15.83&  -15.98 \\
$n_\sat$ (fm$^{-3}$)& 0.159   & 0.159  & 0.162  & 0.160  &   0.160  & 0.158  & 0.158   & 0.161 & 0.160   \\
$K_\sat$ (MeV)      &  239    & 242    & 230    & 240    &   240    & 215    & 241     & 223   & 230     \\
$Q_\sat$ (MeV)      &  -359   & -364   & -405   & -316   &   -350   & -381   & -339    & -393  &  -364  \\
$E_\sym$ (MeV)      &  30.00  & 30.00  & 32.00  & 29.15  &   29.26  & 26.83  & 33.37   & 31.97 &  32.03   \\
$L_\sym$ (MeV)      &  43.9   & 34.9   & 42.4   & 50.9   &   32.4   & 37.6   & 104.3   & 79.1  &  48.3    \\
$K_\sym$ (MeV)      &  -152   & -187   & -113   & -119   &   -191   & -146   & 71      & -43   &  -112 \\
$Q_{\sym}$ (MeV)    & 389     & 462    & 658    & 286    &   440    & 330    & 52      & 131   &  501      \\
\hline\hline
\end{tabular}
\caption{Nuclear empirical parameters for the Skyrme interactions used in the present work.}
\label{tab:EP}
\end{table*}

The general expressions for the pressure \eqref{eq:pum} and the incompressibility modulus \eqref{eq:kea} in AM could be expressed in terms of the parameter $x$ as,
\begin{eqnarray}
P_\um(x,\delta) &=& \frac{n_\sat}{3}(1+3x)^2\frac{\partial e_\um(x,\delta)}{\partial x} \, , \label{eq:pumx}
\\
K_\um(x,\delta) &=& 6(1+3x)\frac{\partial e_\um(x,\delta)}{\partial x}+(1+3x)^2\frac{\partial^2 e_\um(x,\delta)}{\partial x^2} \, . \nonumber \\
\label{eq:kumx}
\end{eqnarray}

Injecting Eq.~\eqref{eq:eAMEP} into the expression for the pressure ~\eqref{eq:pumx}, we obtain
\begin{eqnarray}
P_\um(x,\delta) &=& \frac{n_\sat}{3}\left[ L_\sym \delta^2+ K^p(\delta)x+\frac 1 2 Q^p(\delta)x^2\right]\nonumber \\
&&\hspace{2cm}+ o(x^3)\, ,
\label{eq:pumxs}
\end{eqnarray}
where $K^p=K^\epsilon$ and $Q^p$ reads
\begin{eqnarray}
Q^p(\delta) \equiv Q^p_\sat+Q^p_\sym \delta^2 \, , 
\end{eqnarray}
with
\begin{eqnarray}
Q^p_\sat &\equiv& Q_\sat + 12 K_\sat \, \\
Q^p_\sym &\equiv& Q_\sym + 18 L_\sym + 12 K_\sym \, .
\end{eqnarray}
Note that in finite nuclei, $\vert\delta_A\vert<0.3$ and densities are explored from about $2/3n_\sat$ up to $n_\sat$, which implies $\vert x_A\vert\lesssim 0.1$. In finite nuclei, we could therefore perform an expansion at the same level in $\delta^2$ and in $x$.

The equilibrium density in AM is given by the density for which the mechanical stability is satisfied: $\partial e_\um(x,\delta)/\partial x=0$. From the expression of the pressure~\eqref{eq:pumx} truncated at order $x$, one can deduce in AM~\cite{pie09}, 
\begin{equation}
x_\eq^\um\approx -\frac{L_\sym}{K(\delta)} \delta^2
\approx -\frac{L_\sym}{K_\sat} \delta^2 \, .
\label{eq:xeq}
\end{equation}
The equilibrium density is a function of the isospin asymmetry parameter $\delta$, and it satisfies the limit $n_\eq^\um\rightarrow n_\sat$ for $\delta\rightarrow 0$. At order $\delta^2$ and $x$, one obtains for the equilibrium density $n_\eq^\um$ in asymmetric matter, $n_\eq^\um=n_\sat[1-3(L_\sym/K_\sat)\delta^2]$.

In finite nuclei, the situation is more complex than previously described since: i) the equilibrium density is different from $n_\sat$, due to the finite size terms and ii) the density is not uniform allowing for surface contributions to be sizeable. In uniform matter however, only isospin asymmetry contributes to the shift of the equilibrium density from $n_\sat$, as shown in Eq.~\eqref{eq:xeq}. While neglecting the contribution of the finite-size (FS) terms, expression~\eqref{eq:xeq} provides a good estimation of the average densities in finite nuclei~\cite{Papakonstantinou2013}. In the next section, this density is named $n_\cl$ in the CLDM and we have $n_\cl\approx n_\eq^\um$ for large $A$.

The pressure could be decomposed into a SM and an isospin asymmetry terms: 
\begin{equation}
P_\um=P_\sm + P_\sym \delta^2\, ,
\end{equation}
with
\begin{eqnarray}
P_\sm (n) &=& \frac{n_\sat}{3}\Big[K_\sat x + \frac{1}{2}Q^p_\sat x^2 + ... \Big]    \, ,
\label{eq:pressExpansion}\\
P_{\sym}(n) &=&  \frac{ n_\sat}{3} \Big[ L_\sym  + K^p_\sym x
+ \frac{1}{2}Q^p_\sym x^2 + ...  \Big] \, .
\label{eq:pressSymExpansion}
\end{eqnarray}
We have for instance $P_{\sym}(n_\sat)=n_\sat L_\sym / 3$, as expected.

Similarly, injecting Eq.~\eqref{eq:eAMEP} into \eqref{eq:kumx}, one obtains the following expression for the incompressibility modulus
\begin{eqnarray}
K_\um(x,\delta) &=& K^k(\delta)+Q^k(\delta)x
+\frac 1 2 Z^k(\delta)x^2
+ o(x^3)\, ,
\label{eq:kumxs}
\end{eqnarray}
with $K^k=K^p=K^\epsilon$ and $Q^k=Q^p$, and where the additional coefficient in asymmetric matter reads
\begin{eqnarray}
Z^k(\delta) &\equiv& Z^k_\sat+Z^k_\sym \delta^2 \, , \end{eqnarray}
with
\begin{eqnarray}
Z^k_\sat &\equiv& Z_\sat + 54 K_\sat + 18 Q_\sat \, \\
Z^k_\sym &\equiv& Z_\sym + 54 K_\sym + 18 Q_\sym \, .
\end{eqnarray}
Remark that while $K(\delta)$ controls the isoscalar and isovector dependence of the curvature of the energy per particle in uniform matter ~\eqref{eq:eAMEP}, the incompressibility \eqref{eq:kumx} itself is driven by the parameter $K^k(\delta)$ for $x=0$. The difference between $K(\delta)$ and $K^k(\delta)$ reflects the contribution of the pressure, which is non-zero as soon as the density departs from the equilibrium density $n_\eq^\um$, see Eq.~\eqref{eq:kea}. This contribution is unavoidable, making $K^k(\delta)$ the true isospin dependence of the incompressibility~\cite{pie09}. Fixing $n=n_\sat$ for instance, the parameter which controls the isospin dependence of the incompressibility is $K^k_\sym = K_\sym + 6 L_\sym$, and not K$_\sym$ alone. Considering $L_\sym\approx50$~MeV and $K_\sym\approx -100$~MeV~\cite{Baillot2019,Sagawa2019}, with a lower limit provided by the unitary limit~\cite{Tews2017}, the parameter $K^k_\sym$ is even mostly controlled by $L_\sym$, and only moderately by $K_\sym$. 

In SM the incompressibility modulus can be expressed as a series expansion in $x$ as
\begin{eqnarray}
K_{\sm}(x) &=& K_\sat + (12K_\sat + Q_\sat)x  \nonumber \\
 &+& (27K_\sat+ 9 Q_\sat)x^2 + o(x^3) \, ,
\label{eq:KApowerx}
\end{eqnarray}
and we introduce a new quantity,
\begin{eqnarray}
K_{\sat,\eq} &\equiv& K_\sm(x=x_\eq^\um) \nonumber \\
&=& K_\sat + (12 K_\sat + Q_\sat) x_\eq^\um + o(x^2)\, ,
\end{eqnarray}
which represents the incompressibility modulus of SM for the equilibrium density $n_\eq^\um$:

One can show that in AM the incompressibility modulus at the equilibrium density (\ref{eq:kumxs}) can be expressed as,
\begin{equation}
K_\eq \equiv K_\um(x_\eq^\um,\delta) = K_{\sat,\eq} + K_\sym^k \delta^2 + o(x^2,\delta^4)\, .
\label{eq:keqalt}
\end{equation}
 In Eq.~\eqref{eq:keqalt}, the isovector term $K_\sym^k$=$K_\sym^\epsilon$ (\ref{eq:ksymprime}) depends only on isovector empirical parameters $L_\sym$ and $K_\sym$, while the isoscalar term only depends on isoscalar NEPs $K_\sat$ and $Q_\sat$, provided $x_\eq^\um$ is known (experimentally for instance). In order to perform comparisons with incompressibilities in nuclei $K_A$, it could be relevant to express the incompressibility modulus in AM at equilibrium density $x=x_\eq^\um$ as,
\begin{eqnarray}
K_\eq = K_\sat+K_\tau \delta^2 + o(x^2,\delta^4)\, ,
\label{eq:keq}
\end{eqnarray}
where~\cite{pie09}
\begin{equation}
K_\tau = K_\sym-(6+Q_\sat/K_\sat)L_\sym \, .
\label{eq:ktau}
\end{equation}

We choose nine Skyrme models, BSK14\cite{Goriely07}, BSK16\cite{Chamel08}, F0\cite{Lesinski06}, LNS5\cite{Cao06}, RATP\cite{Rayet82}, SGII\cite{Nguyen81}, SKI2\cite{REINHARD95}, SKO\cite{Reinhard99}, SLy5\cite{Chabanat98}, which NEP are given in Table \ref{tab:EP}. For these nine interactions, while the parameter $K^k_\sym$ is positive for actual values of the NEPs, the parameter $K_\tau$ controlling the isovector dependence of $K_\eq$ is negative since $Q_\sat/K_\sat\approx -1.5$ from Table~\ref{tab:EP}. Note however that the value of $Q_\sat$ has never been measured and its actual value is not necessarily in the range given in Table~\ref{tab:EP}. Aside from the finite size contribution, the ISGMR in finite nuclei is mostly correlated with $K_\eq$, whose isospin dependence is given by $K_\tau$ \cite{pie09}. This is the reason why the isovector dependence of the ISGMR across isotopic chains has been correlated with the parameter $K_\tau$~\cite{Li2010}.

It is clear from the definition of $K_\tau$~\eqref{eq:ktau} that here also, a precise experimental determination of $K_\tau$ does not necessarily lead to a better value for the NEP $K_\sym$, since $K_\tau$ is mostly correlated with $L_\sym$, which is not precisely known. In order to extract $K_\sym$ from experimental investigations, one has to precisely know the values of $L_\sym$ and $Q_\sat$. Note that with about 10\% accuracy~\cite{Khan2012,Khan2013b,Khan2013}, the NEP $K_\sat$ is sufficiently well known in the present case.

\begin{figure}
\centering
\includegraphics[width=0.55\textwidth]{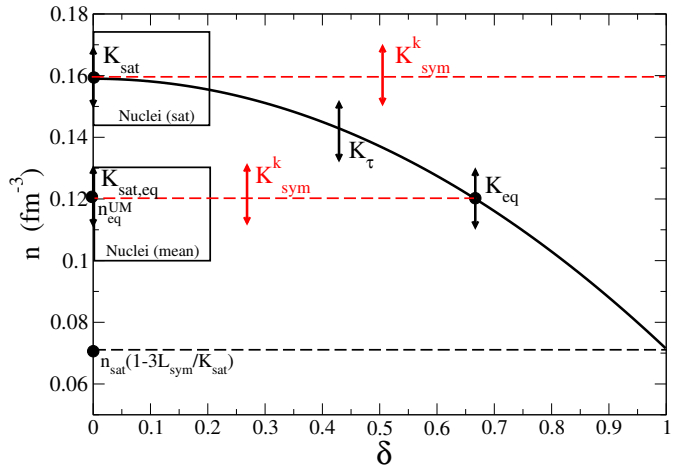}
\caption{Equilibrium points for the BSk12 functional (using its L$_{\sym}$ and K$_{\sat}$ values in Eq. (\ref{eq:xeq})) in the (density, isospin asymmetry) map. The corresponding incompressibilities are schematically indicated. The upper (lower) box are drawn around 
saturation (mean) densities of experimentally accessible nuclei ($\delta<$0.2).}
\label{fig:rhoeq}
\end{figure}

An illustration of the different points where the incompressibility has been introduced is shown in Fig.~\ref{fig:rhoeq}. It displays the behavior of the equilibrium density as a function of $\delta$, on the example of the BSk12 functional. The role of the incompressibility at various densities and isospin is also displayed on the figure. It shows that several incompressibilities at various densities are probed when the GMR is measured in a given nuclei. For instance, in addition to the saturation density, their typical mean density is around 0.11 fm$^{-3}$ \cite{Khan2012,Khan2013b}. It should be noted that $K^k_{\sym}$ drives the isospin dependence of the incompressibility, independently of the considered density, from $x=0$ (saturation point) to $x=x^{\um}_{\eq}$ (equilibrium point).

\subsection{An alternative representation of the nuclear matter energy and incompressibility modulus}
\label{sec:ealt}

In this section, we explore an alternative representation of the uniform matter properties, where the equilibrium density $n_\eq$ is taken, in place of the saturation density $n_\sat$. This alternative representation is equivalent to the existing one up to $\delta^2$, but generates non-quadratic terms. In the view of constraining uniform matter parameters from measurements of incompressibilities in nuclei, it may be more relevant to consider such a representation: the equilibrium density in uniform matter shall be closer -- than the saturation density -- to the average one of the nucleus \cite{Khan2012,Khan2013b}.

In this alternative approach, the associated density parameter is set to be $\xt=(n-n_\eq^\um)/(3n_\eq^\um)$, from which the density $n$ is obtained as $n/n_\eq^\um=1+3\xt$. 

The alternative density parameter $\xt$ can be expressed in term of $x$ as,  
\begin{eqnarray}
\xt = \frac{ x + (L_\sym/K_\sat)\delta^2}{1-3(L_\sym/K_\sat)\delta^2} \, .
\label{eq:xbarx}
\end{eqnarray}

Similarly to Eq.~\eqref{eq:eAMEP} one can expand the energy per particle in term of $\xt$ as
\begin{equation}
e_\um(\xt,\delta) = \tilde{E}(\delta) + \frac 1 2 \tilde{K}(\delta)\xt^2 + \frac 1 6 \tilde{Q}(\delta)\xt^3+\frac 1 {24} \tilde{Z}(\delta)\xt^4+\dots\, ,
\label{eq:eAMEPnew}
\end{equation}
with 
\begin{eqnarray}
\tilde{E}(\delta) &=& e(\xt=0,\delta) = E_\sat+E_\tau\delta^2 \, , \\
\tilde{K}(\delta) &=& \frac{\partial^2 e(\xt,\delta)}{\partial \xt^2}\bat_{\xt=0} = K_\sat+K_\tau\delta^2=K_\eq\, , \label{eq:Keq}\\
\tilde{Q}(\delta) &=& \frac{\partial^3 e(\xt,\delta)}{\partial \xt^3}\bat_{\xt=0} = Q_\sat+Q_\tau\delta^2 \, , \\
\tilde{Z}(\delta) &=& \frac{\partial^4 e(\xt,\delta)}{\partial \xt^4}\bat_{\xt=0} = Z_\sat+Z_\tau\delta^2 \, .
\end{eqnarray}

It should be noted that K$_\tau$ in Eq. (\ref{eq:Keq}) corresponds to (\ref{eq:ktau}), because it is the incompressibility at the equilibrium density, namely $\tilde{K}(\delta)$=K$_{\eq}$.

Imposing the equality between the $\delta^2$ terms in the series expansions ~\eqref{eq:eAMEP} and \eqref{eq:eAMEPnew} orders by orders in $x$, one obtains the following relations,
\begin{eqnarray}
E_\tau&=&E_\sym,\\
K_\tau&=&K_\sym-L_\sym(6K_\sat+Q_\sat)/K_\sat,\\
Q_\tau&=&Q_\sym-L_\sym(9Q_\sat+ Z_\sat)/K_\sat,\\
Z_\tau&=&Z_\sym-L_\sym(12Z_\sat+ Y_\sat)/K_\sat,
\end{eqnarray}
where $Y_\sat$ is the fifth order NEP.  
Note that these equations are a generalization of Eq.~(16) of Ref. \cite{pie09} up to the 4$^{th}$ order, and hence, the above equation for K$_{\tau}$ is the same than the one of the previous subsection.

Eqs.~\eqref{eq:eAMEP} and \eqref{eq:eAMEPnew} are identical up to terms in $\delta^2$. In Eq.~\eqref{eq:eAMEPnew} there are however non-quadratic terms, which are small even when $\delta\sim 1$. The contribution of these non-quadratic terms (because of the denominator in Eq.~\eqref{eq:xbarx}) is even more suppressed by the fact that finite nuclei do not explore large values for $\delta$, since $\vert\delta_A\vert<0.3$, as previously discussed. So it is possible to use both Eq.~\eqref{eq:eAMEP} or Eq.~\eqref{eq:eAMEPnew} to describe the energy in finite nuclei.

Expressing the incompressibility modulus in asymmetric matter~\eqref{eq:kumx} as a function of the density parameter $\xt$:
\begin{equation}
K_{UM}(\xt,\delta) = 6(1+3\xt)\frac{\partial e(\xt,\delta)}{\partial \xt}+(1+3\xt)^2\frac{\partial^2 e(\xt,\delta)}{\partial \xt^2},
\end{equation}
where we have used $(1+3x)\partial/\partial x=(1+3\xt)\partial/\partial \xt$, allows to derive the following expression for the incompressibility: 
\begin{eqnarray}
K_{UM}(\xt,\delta) &=& \tilde{K}(\delta) +[12\tilde{K}(\delta)+\tilde{Q}(\delta)]\xt\nonumber \\
&&\hspace{-0.5cm}
+[27\tilde{K}(\delta)+9\tilde{Q}(\delta)+\frac 1 2 \tilde{Z}(\delta)]\xt^2
+ o(\xt^3)\, .
\label{eq:kgen}
\end{eqnarray}
Eq.~\eqref{eq:kgen} provides a series expansion of the incompressibility modulus in asymmetric matter up to $\xt^2$ and $\delta^2$, which is convenient to use when constraining the uniform matter incompressibility from measurements in nuclei. 

We will use the alternate representation developed in this section and confront it with the standard expansion of the nuclear matter energy in Sec.~\ref{sec:sound_speed} where we present our analysis of the speed of sound.

\subsection{Constraints on $K_\sym$}

From the existence of a lower bound on the energy of NM, on the basis of unitary-gas considerations, the following constraint on $K_\sym$ was obtained~\cite{Tews2017},
\begin{equation}
K_\sym\approx -306.0 + 3.41L_\sym \pm 28.3\hbox{ MeV }\, ,
\label{eq:ksymunit}
\end{equation}
when models with $K_\sat >275$~MeV are excluded. Considering $L_\sym \approx 50$~MeV for instance, this constraint imposes $K_\sym \geqslant \simeq -150$~MeV (see Table \ref{tab:EP}).

In Eq.~\eqref{eq:ksymunit}, the coefficients of the correlation are obtained from a fit to a given set of model realizations. In the following, we will demonstrate the existence of a lower limit from purely theoretical considerations. 

It is possible to express the equilibrium density from $K_\sym$, by solving the mechanical stability condition $\partial e_\um(x,\delta)/\partial x=0$, with an expansion of the energy to $x^3$ and beyond $\delta^2$ approximation. The physical solution of this second order equation is
\begin{eqnarray}
x_{\eq,2}^\um(\delta) = \frac{K(\delta)}{Q(\delta)}\left[-1+\sqrt{1-2\frac{L_{\sym} Q(\delta)}{K(\delta)^2}\delta^2}\right] \, ,
\label{eq:xeq2}
\end{eqnarray}
satisfying the limit $n_{\eq,2}^\um\rightarrow n_\sat$ as $\delta\rightarrow 0$. Eq.~\eqref{eq:xeq2} is well defined if $K(\delta)^2\geq2L_{\sym}Q(\delta)\delta^2$ for all values of $\delta$ for which equilibrium density is defined, which is ranging from SM to very asymmetric matter. There is no equilibrium density in NM, but there is still an equilibrium very close to NM. Since Eq.~\eqref{eq:xeq2} weakly depends on $\delta$ for isospin asymmetries close to NM, we fix $\delta=1$ in Eq.~\eqref{eq:xeq2} for simplicity. We then obtain: $K_\sym\geq-K_\sat+\sqrt{2L_\sym Q(\delta=1)}$ or $K_\sym \leq-K_\sat-\sqrt{2L_\sym Q(\delta=1)}$. Considering typical values for the NEPs extracted from Table~\ref{tab:EP}, $\sqrt{2L_\sym Q(\delta=1)}\sim 70-100$~MeV, so the previous condition gives $K_\sym\gtrsim-150$~MeV or $K_\sym \lesssim-350$~MeV. Since the second case is excluded by the constraint on $K_\sym$ given by considerations based on the unitary-gas~\cite{Tews2017}, we are then left with the first condition alone: 
\begin{equation}
K_\sym\geq-K_\sat+\sqrt{2L_\sym Q(\delta=1)}\sim-150\hbox{ MeV}\, .
\label{eq:ksymvalue}
\end{equation}

Note that using the model averaged values of $Q_\sat$ and $Q_\sym$~\cite{meta1}, we have the following condition: $Q(\delta)> 0$ for all $\delta$, constraining $Q_\sat$ and $Q_\sym$, as $Q_\sym>-Q_\sat$. However, this relation is not always satisfied, as shown in table~\ref{tab:EP}.

Other estimates of $K_\sym$ from neutron stars observations have been suggested: from X-ray thermal emission on seven LMXBs, it was found $K_\sym=-85^{+82}_{-70}$~MeV~\cite{Baillot2019};from the analysis of GW170817 it was determined that $-259<K_\sym<32$~MeV
\cite{Carson2019}.

Using recent FRDM mass model\cite{frdm2012} and the neutron skin of $^{48}$Ca extracted from ($p,p^\prime$) experiments and fixing the nuclear incompressibiliity $K_\sat=225\pm20$~MeV from up-to-date experimental data of ISGMR of $^{208}$Pb, it was found that $K_\sym=-120\pm40$~MeV~\cite{Sagawa2019}. The constraint obtained from FRDM mass model leads to fix $E_\sym=32.3\pm0.5$~MeV and $L_\sym=53.5\pm15$~MeV. The neutron skin experiment gives $L_\sym=42\pm15$~MeV. 

A compilation of 16 results from independent analyses of neutron
star observational data since GW170817 lead to the following expectation $K_\sym \approx - 107 \pm 88$~MeV~\cite{BALi2021}. All these data tend to point towards negative values of $K_\sym$, with a centroid located around $-100$~MeV. The uncertainty is difficult to estimate, but a conservative value may be around $-100$~MeV. Note that these results are compatible with the constraint~\eqref{eq:ksymvalue} that we derived.

\section{The compressible liquid drop model with a density dependence of the surface tension}

The CLDM has been originally developed on top of the liquid drop model, where the bulk term is a constant~\cite{Myers69}. In the CLDM~\cite{Myers69,Weiss69}, the bulk term is density-dependent and the density is fixed variationally  by the mechanical stability condition. In the present approach, we suggest an extension of the CLDM by introducing a density-dependent surface term. We show that the present eCLDM could describe accurately both the energy of finite nuclei in their ground state as well as the ISGMR energy.

\subsection{Density dependent surface tension}

The novelty of the present work is the introduction of a density dependent surface tension, which is expressed as,
\begin{eqnarray}
 \sigma_\surf (n_\cl, I_\cl) = \sigma_\surf(I_\cl)\left[1 + a_\surf f(A_\cl) x_\cl^2\right]\, ,
  \label{eq:sigmaeCLDM}
\end{eqnarray}
where $x_\cl = (n_\cl - n_\sat)/3n_\sat$ and the parameter $a_\surf$ controls the density dependence of the surface energy. In practice, it encodes the deviation from $n_\sat$. It is then larger for nuclei for which $n_\cl$ is farther from $n_\sat$, i.e. for light and intermediate mass nuclei as well as for exotic nuclei. In Sec.~\ref{sec:KA} we suggest a way to estimate $a_\surf$ by using a single microscopic calculation of $K_A$ in $^{100}$Sn.

In Eq.~\eqref{eq:sigmaeCLDM}, the function $f(A_\cl)$ is defined as,
\begin{equation}
f(A_\cl) = \frac{1}{1 + \exp[-(A_\cl - A_0)/A_w]},
\end{equation}
where $A_\cl$ is the mass number of the considered nucleus. This function has been introduced to suppress the density dependence of the surface tension in light nuclei, where it appears to be unrealistically too large. From a qualitative study, we suggest the following values for the parameters of the function $f$: $A_0 = 70$ and $A_w = 10$. 

\begin{figure}[t]
\centering
\includegraphics[scale=0.32]{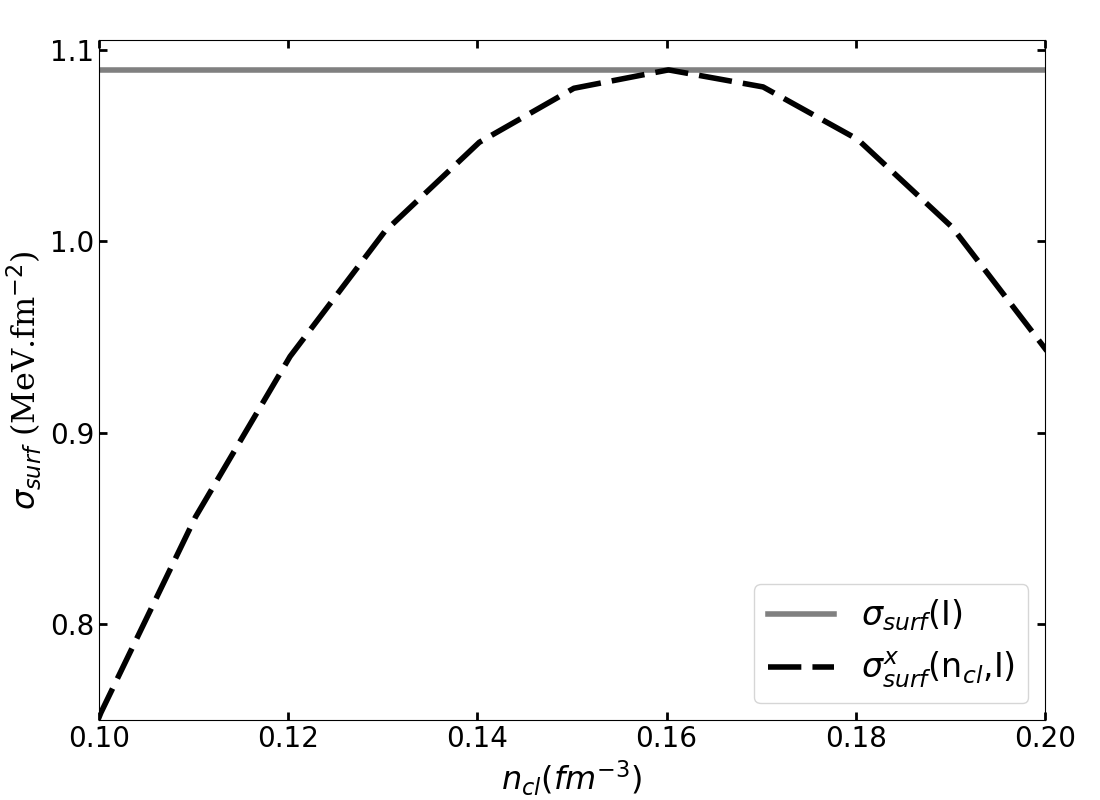}
\caption{Surface tension for $^{120}$Sn. Continuous gray line show the result without density dependence on $\sigma_\surf$. Dashed line shows the result with density-dependent surface tension given by Eq.~\eqref{eq:sigmaeCLDM}.}
\label{fig:sigSurf}
\end{figure}

Figure \ref{fig:sigSurf} shows a comparison of the surface tension $\sigma_\surf$ from the CLDM and eCLDM as function of the cluster density $n_\cl$. The figure shows a bell shape for the eCLDM due to its quadratic dependence on $x_\cl$, in contrast with the horizontal line of the CLDM which does not depend on the density. The eCLDM simulates a decrease of the surface tension by about 30\% from $n_\cl\approx 0.16$~fm$^{-3}$ down to $n_\cl\approx 0.10$~fm$^{-3}$. So for typical values of the cluster density, eCLDM reduces the surface tension to a large amount.

Note that we have also investigated other functionals of the density. For instance, we have studied a correction term similar to Eq.~\eqref{eq:sigmaeCLDM}, replacing $x_\cl$ by $n_\cl$ and fixing $a_\surf$ as described in Sec.~\ref{sec:KA}. We found that this correction changes the pressure to a large amount, shifting up the cluster density $n_\cl$ to unrealistic values (above $0.25$~fm$^{-3}$ in some cases).

It should be noted that we have chosen the exponent of the density-dependent term in Eq.~\eqref{eq:sigmaeCLDM} to be two. The reason is twofold: first, it approximately satisfies the stationarity of the surface tension w.r.t the density, see Ref.~\cite{Blaizot80} for more details, and second, with such a power, it directly contributes to the incompressibility modulus in finite nuclei. Note that a correction proportional to $x_\cl$ has been suggested in Ref.~\cite{Iida04}, and analysed in view of its impact on the neutron skin. However, such a term does not satisfies the requested stationary of the surface tension and does not contribute to the incompressibility in finite nuclei. 

\subsection{Incompressibility in finite nuclei: $K_A$}
\label{sec:KA}

The incompressibility $K_A$ in finite nuclei is defined as,
\begin{eqnarray} 
K_A &\equiv&  9 n_\cl\frac{\partial^2  \epsilon_{A}}{\partial n_{\cl}^2}\Bigr|_{A}, 
\label{eq:KA1}
\end{eqnarray}
with the energy density given by $\epsilon_{A} = e_A n_\cl$. 

According to Eq.~\eqref{eq:KA1} by deriving twice the energy density w.r.t the cluster density, we obtain the incompressibility in a nucleus as,
\begin{widetext}
\begin{eqnarray}
K_{A} &=& K_{\sat} + K_{\tau}\delta^2  +\mathcal{C}_\coul \frac{3}{5}\frac{e^2}{r_0} \left( 8 +\frac{Q_{\sat}}{K_{\sat}} \right) Z^2A^{-4/3} \nonumber \\
&+& \mathcal{C}_\surf\Bigg[ 8 \pi r^2_{\cl} \sigma_\surf\left(11 +  \frac{Q_{\sat}}{K_{\sat}} \right) -12\pi n_\cl r_\cl^2 \frac{\partial \sigma_\surf }{\partial n_\cl}\left( 10 +   \frac{Q_{\sat}}{K_{\sat}} \right) 
+ 36\pi n_\cl^2 r_\cl^2 \frac{\partial^2 \sigma_\surf }{\partial n_\cl^2} \Biggr] A^{-1/3} \, .
\label{eq:kacomplete}
\end{eqnarray}
\end{widetext}
where $\mathcal{C}_\coul$ and $\mathcal{C}_\surf$ are coefficients (close to 1) optimized in order to reproduce nuclear experimental masses. A detailed derivation of $K_A$  is given in App.~\ref{app:KAderivation}. The values of the parameters used in the present work are given in Tabel \ref{tab:FS}. We can identify in the above expression the incompressibility modulus $K_\sat$, the isospin term $K_\tau$, the Coulomb and surface terms respectively. We have arranged this expression to be comparable with Eq.~(6.3) of Blaizot~\cite{Blaizot80}. Note that the terms in the surface contribution which are proportional to the derivative of the surface tension w.r.t. the cluster density are absent in usual CLDM while in the eCLDM, these terms become proportional to the constant $a_\surf$ introduced in Eq.~\eqref{eq:sigmaeCLDM}.

\subsection{Definition of the parameter $a_\surf$ and incompressibility predictions within the eCLDM}

The new parameter $a_\surf$ controlling the density dependence of the surface tension, is fixed to reproduce the microscopic prediction for the incompressibility $K_A$ in the doubly magic $N=Z$ nucleus $^{100}$Sn. The values $a_\surf$ and the microscopic prediction from constrained Hartree-Fock-Bogoliubov (CHFB), $K_{A,CHFB}(^{100}\hbox{Sn})$, are shown in Table \ref{table:parameterC} for the nine Skyrme interactions. The accuracy with which the microscopic prediction is reproduced by the eCLDM is fixed to $< 1$~MeV. 

\begin{table}[b]
\centering
\tabcolsep=0.7cm
\def\arraystretch{1.5}
\begin{tabular}{ccc}
\hline\hline
       & $K_{A,CHFB}(^{100}\hbox{Sn})$ & $a_\surf$ \\
       & MeV &  \\\hline
 BSK14\cite{Goriely07} & 153.6  & -19.95 \\
 BSK16\cite{Chamel08} & 154.4  & -20.00 \\
   F0\cite{Lesinski06}  & 142.3  & -19.90 \\
 LNS5\cite{Cao06}  & 150.7  & -20.95 \\
 RATP\cite{Rayet82}  & 147.9  & -20.85 \\
 SGII\cite{Nguyen81}  & 133.2  & -19.55 \\
  SKI2\cite{REINHARD95} & 155.2  & -20.00 \\
  SKO\cite{Reinhard99}  & 139.3  & -19.55 \\
 SLy5\cite{Chabanat98}  & 142.8  & -20.05 \\
\hline\hline
\end{tabular}
 \caption{For a set of Skyrme interactions, microscopic Constrained Hartree-Fock-Bogoliubov predictions for $K_A$ in $^{100}$Sn used in the calibration of the parameter $a_\surf$.}
\label{table:parameterC}
\end{table}

Since the parameter $a_\surf$ is found to be very stable and close to $\sim - 20$, the fit of the eCLDM is made into two steps: First, the values of the coefficients $\mathcal{C}_{\surf,\sat}$, $\mathcal{C}_{\surf,\sym}$, and $\mathcal{C}_\coul$ are fitted to better reproduce the experimental nuclear masses, using an initial value $a_\surf = -20$ (see Table~\ref{tab:FS}), then in a second step, the value of $a_\surf$ is accurately fixed by fitting $K_{A,CHFB}(^{100}\hbox{Sn})$ for each of the Skyrme model (see Table~\ref{table:parameterC}). For details about the microscopic CHFB approach, we refer for instance to Ref.~\cite{Khan2013}.

\begin{figure}[t]
\centering
\includegraphics[scale=0.34]{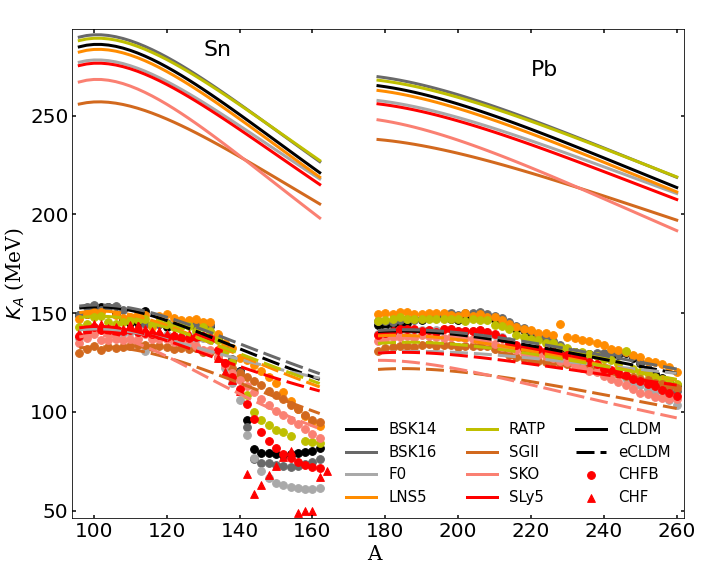}
\caption{Comparison of $K_A$ for Pb an Sn isotopic chains. Continuous (dashed) lines show results with CLDM (eCLDM). Constrained Hartree-Fock-Bogoliubov (CHFB) calculations are shown in dots. Red triangles shows CHF calculations for SLy5, i.e., microscopic calculations for Sn isotopes without pairing.}
\label{fig:KA_PbSn-SLy5DDME2}
\end{figure}

In Fig.~\ref{fig:KA_PbSn-SLy5DDME2}, we show as function of $A$, for Sn and Pb isotopes, the comparison of the CLDM (solid lines) and eCLDM \eqref{eq:kacomplete} (dashed lines) predictions against the microscopic predictions (circles) for $K_A$, based on the constrained Hartree-Fock-Bogoliubov (CHFB) approach, for the set of Skyrme interaction listed in Tab.~\ref{table:parameterC} (see Refs.~\cite{Khan2012,Khan2013b} for more details on the microscopic CHFB approach). We use the microscopic radii calculated by each interaction, to transform $E_\gmr$ into $K_A$ using Eq.~(\ref{eq:egmr}). For SLy5, our results are identical to the ones given in Ref.~\cite{Khan2013}. By comparing CHF (red triangles) and CHFB (red circles), we see that pairing contributes to reduce the shell effects around $A\approx 140$ in Sn and makes the isotopic evolution of $K_A$ smoother.

As stated by Blaizot~\cite{Blaizot80}, the CLDM predictions are largely overestimating $K_A$, since the surface energy is not explicitly density-dependent. By adding the new term~\eqref{eq:sigmaeCLDM} for the surface energy (Eq.~\eqref{eq:sigma_old}), the eCLDM reproduces the isotopic dependence of $K_A$ as predicted by the microscopic CHFB approach (note that only one nucleus ($^{100}$Sn) has been used for the calibration of $a_\surf$). 

In Sn isotopes, one can note a marked step for $K_A$ for $A\gtrsim 132$, in microscopic predictions, which is not present in the eCLDM prediction. The eCLDM predicts instead a continuous decrease of $K_A$ over the isotopic chain. A similar feature, while not as pronounced, is observed for Pb isotopes for $A\gtrsim 208$. Microscopically, these steps are understood as originating from shell effects: in $^{132}$Sn, they are 12 occupied 1h$_{11/2}$ states below the Fermi level and 10 unoccupied 1h$_{9/2}$ states. For $^{208}$Pb, the Fermi level is for $1i_{13/2}$ and the next orbital is $1i_{11/2}$. The $\Delta L=0$ isoscalar oscillation is therefore enhanced for neutron-rich systems belonging to these two isotopic chains. Since shell effects are not present in the eCLDM, such steps could not be described by our macroscopic approach. The decrease of $K_A$ as nuclei get more and more neutron-rich is however well reproduced by the eCLDM approach. Such a dependence on $A$ depends on the choice of the NEP, as illustrated in the appendix section \ref{sec:sensitivity}.

\begin{table*}[t]
\centering
\tabcolsep=0.26cm
\def\arraystretch{1.5}
\begin{tabular}{cccccccl}
\hline\hline
A & Z &   $e_A$ & $c_A$  &  $K_{A}$ & $\bar{K}_\surf$ & $\bar{K}_\coul$ &  $E_{\gmr}$  \\
 &  &   (MeV) &  & (MeV) & (MeV)  & (MeV) & (MeV) \\
\hline\hline
100  &  50  &  -8.11 / -8.08 [-8.25] &  -1.2 / 1.5  &  143.2 / 278.3  &  -276.7 / 347.7  &  -5.0 / -4.9  &  19.5 / 26.6  \\ 
106  &  50  &  -8.40 / -8.38 [-8.43] &  -1.3 / 1.5  &  142.7 / 277.3  &  -289.6 / 345.2  &  -5.0 / -4.9  &  19.1 / 26.0  \\ 
114  &  50  &  -8.55 / -8.53 [-8.52] &  -1.3 / 1.5  &  139.5 / 272.2  &  -305.1 / 336.5  &  -5.0 / -4.9  &  18.3 / 25.1   [15.9]  \\ 
120  &  50  &  -8.53 / -8.52 [-8.50] &  -1.4 / 1.4  &  136.1 / 266.6  &  -315.0 / 327.4  &  -5.0 / -4.9  &  17.6 / 24.4  [15.5]  \\ 
180  &  82  &  -7.72 / -7.72 [-7.73] &  -1.6 / 1.5  &  130.0 / 256.8  &  -366.7 / 348.2  &  -4.9 / -4.9  &  14.9 / 20.8  \\ 
200  &  82  &  -7.87 / -7.88 [-7.88] &  -1.6 / 1.4  &  128.6 / 247.9  &  -367.4 / 329.7  &  -4.9 / -4.9  &  14.2 / 19.6  \\ 
208  &  82  &  -7.83 / -7.84  [-7.87]&  -1.6 / 1.4  &  127.1 / 243.1  &  -367.0 / 320.5  &  -4.8 / -4.8  &  13.8 / 19.1  [13.5] \\ 
   \hline\hline
\end{tabular}
\caption{Binding energies, ratio $c_A = \bar{K}_{A,\surf}/K_\sat$, incompressibility $K_A$ in finite nuclei and the contributions from the surface and Coulomb terms, and the ISGMR energies for a set of Sn and Pb isotopes, using the meta-model version of the SLy5 nuclear interaction. On the two sides of the bar $/$ are compared the values obtained from the eCLDM and the CLDM approaches. The experimental values for the binding energies and for the $E_{\rm ISGMR}$, when available, given by the 2020 Atomic Mass Evaluation (AME) table~\cite{AME2020} and by Garg et al\cite{GARG2018}, respectively, are given inside brackets. }
\label{table:EnerKASLy5}
\end{table*}

As a first application of the eCLDM, we compute the binding energies, the incompressibilities $K_A$ and ISGMR energies for several Sn and Pb isotopes and for the SLy5 Skyrme interaction (see Tab.~\ref{table:EnerKASLy5}). All these quantities are given for both the eCLDM and the CLDM approaches. Experimental data for the binding energies from AME2020 table~\cite{AME2020} are also given. We have also calculated the ratio $c_A= K_{A,\surf}/K_\sat$ for the set of nuclei. Interestingly, we found $c_A \approx - 1.3$, which is compatible with the calculations of Ref.~\cite{Patra2002} deduced from a microscopic approach. Consistently with Fig.~\ref{fig:KA_PbSn-SLy5DDME2}, the value obtained for $K_A$ with the eCLDM is considerably reduced, compared to the one provided from the CLDM, illustrating the impact of the density-dependent surface energy term. $K_\surf$ is also given in Tab.~\ref{table:EnerKASLy5}. The contribution of the density-dependent surface energy term is large: it changes the sign of the term $K_\surf$, from positive (CLDM) to negative (eCLDM). The $A$ dependence of $K_\surf$ is also strongly modified with the density-dependent surface energy term. The value for $K_\coul$ is not much impacted by the density-dependent surface energy term. In addition, $K_\coul$ is compatible with the value extracted from the liquid drop expansion~\cite{Sagawa2007} and are rather insensitive to the nuclear interaction. Finally, we show, in the last column, the values for ISGMR energies. Since this values are directly impacted by $K_A$, see Eq.~\eqref{eq:egmr}, the eCLDM shows a reduction for the $E_{\rm ISGMR}$ energies. Note that this reduction makes the eCLDM results closer to the experimental values.

Tab.~\ref{table:EnerKASLy5} illustrates one of the main feature of the eCLDM approach: the present density-dependent surface energy term has a small impact on the binding energies, but a large contribution to the incompressibility modulus $K_A$ in finite nuclei. This justifies our fitting protocol previously described. It also shows that the low order NEP could be adjusted to the nuclear mass table quite independently to the higher order NEP which are fitted to $K_A$.

\section{Confrontation to the nuclear experimental data}

In this section, we confront the eCLDM to the nuclear data. To do so, we first list the experimental data used for the analysis. By using the Markov chain Monte Carlo (MCMC) approach, we then vary a set of NEP all together in order to extract the best parameters set reproducing the experimental data. 
A sensitivity analysis is shown in the appendix~\ref{sec:sensitivity} where we illustrate, in a complementary way, the individual influence of the NEP to the prediction of $K_A$.

\subsection{Experimental data for $K_A$}

\begin{table}
\begin{ruledtabular}
\begin{tabular}{lllll}
 & $E_\gmr$ & $E_\gmr$ & $R_{A}$ & $K_A$ \\
 & (MeV) & (MeV) & (fm) & (MeV) \\
 & from Ref.~\cite{GARG2018} & (this work) & (SLy5) & from Eq.~\eqref{eq:egmr} \\
\hline
$^{90}$Zr & 17.58$^{+0.06}_{-0.04}$ & $17.62\pm{0.07}$ & 4.256 & $135.6\pm{1.1}$ \\
 & 17.66$^{+0.07}_{-0.07}$ & \\
$^{92}$Zr & 17.71$^{+0.09}_{-0.07}$ & $17.62\pm{0.12}$  & 4.293 & $138.0\pm{1.9}$ \\
 & 17.52$^{+0.04}_{-0.04}$ & \\
$^{94}$Zr  & 15.75$^{+0.27}_{-0.15}$ & $15.80\pm{0.21}$ & 4.330 & $112.9\pm{3.0}$ \\
$^{112}$Sn & 15.23$^{+0.26}_{-0.14}$ & $15.69\pm{0.44}$ & 4.556 & $123.2\pm{6.9}$ \\
& 16.10$^{+0.10}_{-0.10}$ & \\
$^{114}$Sn & 15.90$^{+0.10}_{-0.10}$ & $15.90\pm{0.10}$ &  $4.585$ & $128.2\pm{1.6}$ \\
$^{116}$Sn & 15.70$^{+0.10}_{-0.10}$ & $15.70\pm{0.10}$ &  4.614 & $126.5\pm{1.6}$ \\
$^{118}$Sn & 15.60$^{+0.10}_{-0.10}$ & $15.60\pm{0.10}$ & 4.641 & $126.4\pm{1.6}$ \\
$^{120}$Sn & 15.50$^{+0.10}_{-0.10}$ & $15.50\pm{0.10}$ & 4.667 & $126.2\pm{1.6}$ \\
$^{122}$Sn & 15.20$^{+0.10}_{-0.10}$ & $15.20\pm{0.10}$ & 4.691 & $122.6\pm{1.6}$ \\
$^{124}$Sn & 14.33$^{+0.17}_{-0.14}$ & $14.72\pm{0.40}$ & 4.715 & $116.2\pm{6.3}$ \\
& 15.10$^{+0.10}_{-0.10}$ & \\
$^{132}$Sn$^\dagger$ & 14.80 & 14.80 &  4.803 & 121.8 \\
$^{204}$Pb & 13.70$^{+0.10}_{-0.10}$ & $13.70\pm{0.10}$ & 5.516 & $137.7\pm{2.0}$ \\
$^{206}$Pb & 13.60$^{+0.10}_{-0.10}$ & $13.60\pm{0.10}$ & 5.532 & $136.5\pm{2.0}$ \\
$^{208}$Pb & 13.50$^{+0.10}_{-0.10}$ & $13.50\pm{0.10}$ & 5.548 & $135.3\pm{2.0}$ \\
\end{tabular}
\end{ruledtabular}
$^\dagger$Fictitious data.
\caption{Experimental data for $E_\gmr$ and $K_A$, considered in this work. 
}
\label{tab:KA}
\end{table}

We aim at reproducing together the values of $K_A$ in $^{90,92}$Zr,  $^{112-124}$Sn, $^{204-208}$Pb from Ref.~\cite{GARG2018}, see Tab.~\ref{tab:KA} for detailed values. We do not consider here the experimental GMR energy measured for $^{94}$Zr and reported in Ref.~\cite{GARG2018}, since it is very different from the one measured in $^{90}$Zr and $^{92}$Zr. It is not possible, for our modeling to reproduce this data, as it is shown hereafter in Fig.~\ref{fig:mcmc_KA}. In addition, we investigate the role of a fictitious measurement of the GMR energy in $^{132}$Sn and explore possible consequences for the determination of NEP. 

We first report, in Tab.~\ref{tab:KA}, the experimental data listed in Ref.~\cite{GARG2018}. For some nuclei there are different values obtained from different experiments, see for instance $^{90}$Zr, $^{92}$Zr, $^{112}$Sn and $^{124}$Sn (the largest differences between different experimental measurements are for $^{112}$Sn and $^{124}$Sn). In the following, we adopt an agnostic approach w.r.t. these data and we then equally treat the measurements. It should be noted that we have then re-calculated averaged centroids and standard deviations for nuclei were two experimental values are reported, generating a new distribution summing the individual ones. We have then determined the value for $K_A$ using Eq.~\eqref{eq:egmr}, where the total radius $R_{A}$ is provided by an CHFB calculation~\cite{Bennaceur2005} using SLy5~\cite{Chabanat98} Skyrme interaction. The last column in Tab.~\ref{tab:KA} gives the experimental values for $K_A$ which are used in the confrontation of our eCLDM to nuclear data.

\subsection{Best parameter set from Markov-chain Monte Carlo approach}
\label{sec:mcmc}

In this subsection, we vary a set of NEP in order to determine the best parameters reproducing the experimental data. We first present the experimental data and then the Markov-chain Monte Carlo (MCMC) approach we adopt.

The confrontation between the experimental data for the incompressibility $K_A$ (see tab.~\ref{tab:KA}), and the model predictions, is based on the loss functions $\chi_{K_A}$, which is defined as
\begin{equation}
\chi_{K_A}^2 = \frac{1}{N_{K_A}} \sum_i \left(\frac{K_{A,i}^{\exp}-K_{A,i}^\ecldm}{\delta K_{A,i}^{\exp}}\right)^2,
\end{equation}
where $i$ runs over the following isotopes: $^{90,92}$Zr, $^{112-124}$Sn, $^{204-208}$Pb. We also explore a fictitious data for $^{132}$Sn, since and experimental value of the GMR centroid is currently under analysis
\cite{RikenReport17}.

\begin{table*}[t]
\begin{ruledtabular}
\centering
\tabcolsep=0.1cm
\def\arraystretch{1.5}
\begin{tabular}{ccccccccccc}
& E$_{\sat}$ & n$_{\sat}$  & K$_{\sat}$ & Q$_{\sat}$ & Z$_{\sat}$ & E$_{\sym}$ & L$_{\sym}$ & K$_{\sym}$ & Q$_{\sym}$ & Z$_{\sym}$ \\
& (MeV) & (fm$^{-3}$) & (MeV) & (MeV) & (MeV) & (MeV) & (MeV) & (MeV) & (MeV) & (MeV) \\
\hline
From Ref.~\cite{meta1} & -15.8$\pm$0.3 & 0.155$\pm$0.005 & 230$\pm$20 & 300$\pm$400 & -500$\pm$1000 & 32$\pm$ 2 & 60$\pm$15 & -100$\pm$100 & 0$\pm$ 400 & -500$\pm$1000\\
dist1f and dist2f & -15.8    & 0.155 & [210,250] & [-1800,600] & -500 & 32 & [40,60] & [-300,100] & 0 & -500 \\
dist3f & -15.8    & 0.155 & [210,250] & [-1800,600] & -500 & 32 & [80,100] & [-300,100] & 0 & -500
\end{tabular}
\end{ruledtabular}
\caption{Priors for the NEP from Ref.~\cite{meta1} (first raw), and priors considered in the present analysis to set-up dist1f, dist2f and dist3f. The values given in interval $[a,b]$ imply that a flat prior is considered in the MCMC approach. Other NEP are fixed to the indicated values. We considered also the following parameters: $M^*_\sat=0.7$, $\Delta M^*_\sat=-0.1$, $b_\sat=6.9$ and $b_\sym=0$.}
\label{tab:NEP}
\end{table*}

The eCLDM is also fine-tuned to experimental nuclear masses. The associated loss function $\chi_{E}$ is defined as
\begin{equation}
\chi_{E}^2 = \frac{1}{N_{E}} \sum_i \left(\frac{E_{i}^{\exp}-E_{i}^\ecldm}{\delta E_{i}^{\exp}}\right)^2\, ,
\end{equation}
where $i$ runs over a subset of experimental binding energy extracted from the 2020 AME mass table~\cite{AME2020}. To speed-up the computing time, we do not consider all nuclei in the mass table, as in Ref.~\cite{Grams2022a} for instance, but instead we confront the mass model to a subset of it. To do so, we picked-up one out of hundred data. We have checked that this selection does not impact our results, as discussed below. 

In the following we fix the NEP $E_\sat$, $E_\sym$, and $n_\sat$ to their empirical expectations, as reported in Tab.~\ref{tab:NEP}. We vary the other NEP, $K_\sat$, $Q_\sat$, $L_\sym$ and $K_\sym$, considering flat priors inside the boundaries given in Tab.~\ref{tab:NEP} and defining the prior loss function $\chi_{prior}$. The higher order NEP $Q_\sym$, $Z_\sat$ and $Z_\sym$ have no impact on the present analysis. Hence, they are fixed to values determined from analyses of model predictions, see Ref.~\cite{Margueron2018a}. Their value is also given in Tab.~\ref{tab:NEP}. Finally, the effective mass, which is parameterized by $M^*_\sat$ and $\Delta M^*_\sat$, is also fixed in the present study.

The total loss function is obtained as the sum of $\chi_{K_A}$, $\chi_{E}$, and $\chi_{prior}$. We explore three scenarios in the present study: 
\begin{itemize}
\item dist1 \& dist1f: all known experimental data are considered for $K_A$ ($^{90,92}$Zr, $^{112-124}$Sn and $^{204-208}$Pb) and the priors are taken flat, as given in Tab.~\ref{tab:NEP}. 
\item dist2 \& dist2f: same as dist1 \& dist1f but considering a fictitious value for $K_A$ in $^{132}$Sn, as given in in Tab.~\ref{tab:KA}.
\item dist3 \& dist3f: same as dist2 \& dist2f but considering a large prior for $L_\sym$, as given in Tab.~\ref{tab:NEP}.
\end{itemize}
The difference between the cases dist$i$ and dist$i$f ($i=1$, 2, 3) are that dist$i$f includes the fine tuning of the eCLDM to the experimental nuclear masses while dist$i$ does not. In the following results, we observe that there are very little differences between dist$i$ and dist$i$f, since the NEPs ($E_\sat$, $E_\sym$ and $n_\sat$) which play a major role in the determination of the nuclear masses are not varied in the present study.

\begin{figure}[t]
\centering
\includegraphics[scale=0.35]{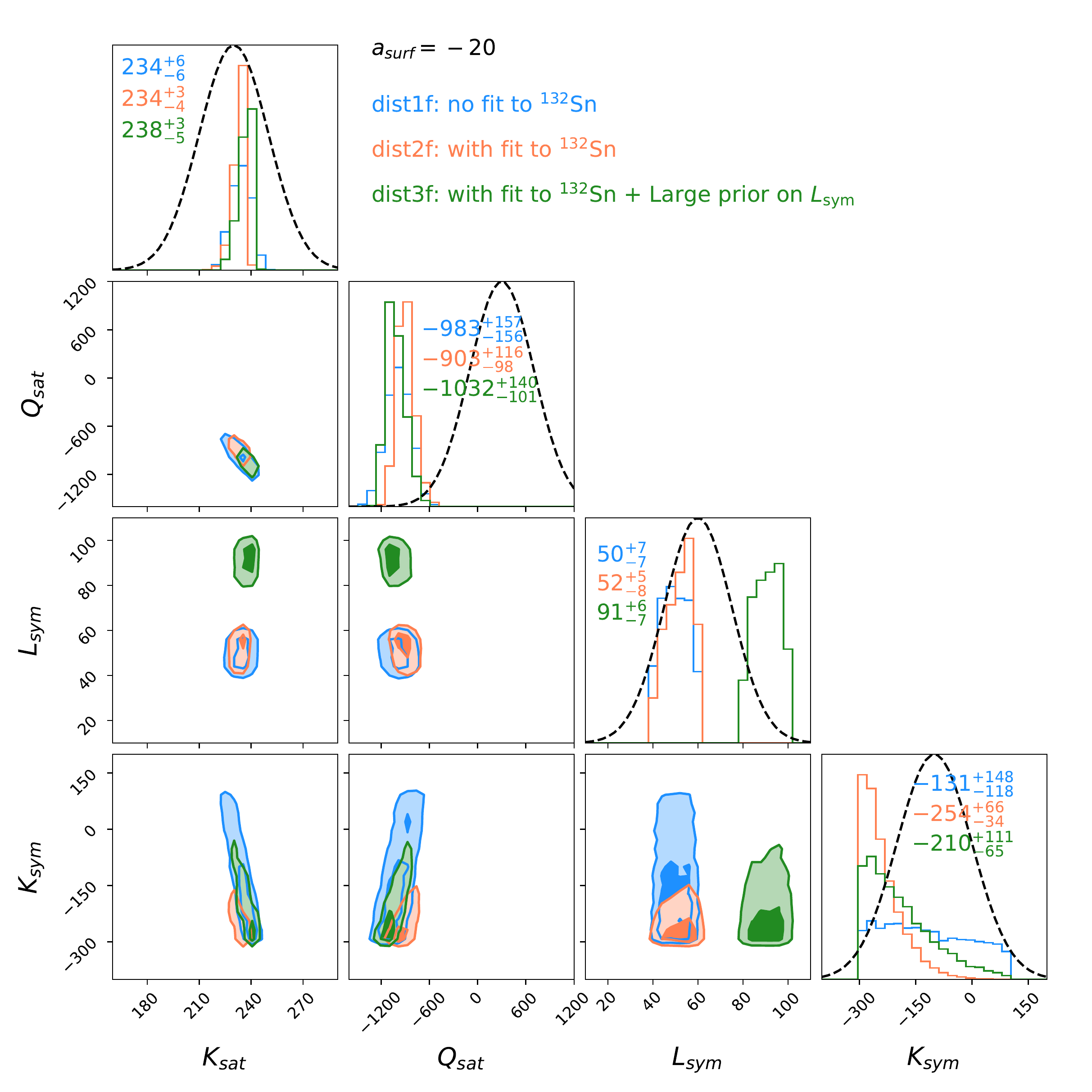}
\caption{Marginalized distributions for the NEP parameters in the three cases which are considered here: dist1f (considering all experimental data), dist2f (adding a fictitious measurement for $^{132}$Sn) and dist3f (with a prior for $L_\sym=90\pm10$~MeV). Note that for dist1f and dist2f the prior is $L_\sym=50\pm10$~MeV.}
\label{fig:mcmc_corner}
\end{figure}

\begin{figure}[t]
\centering
\includegraphics[scale=0.5]{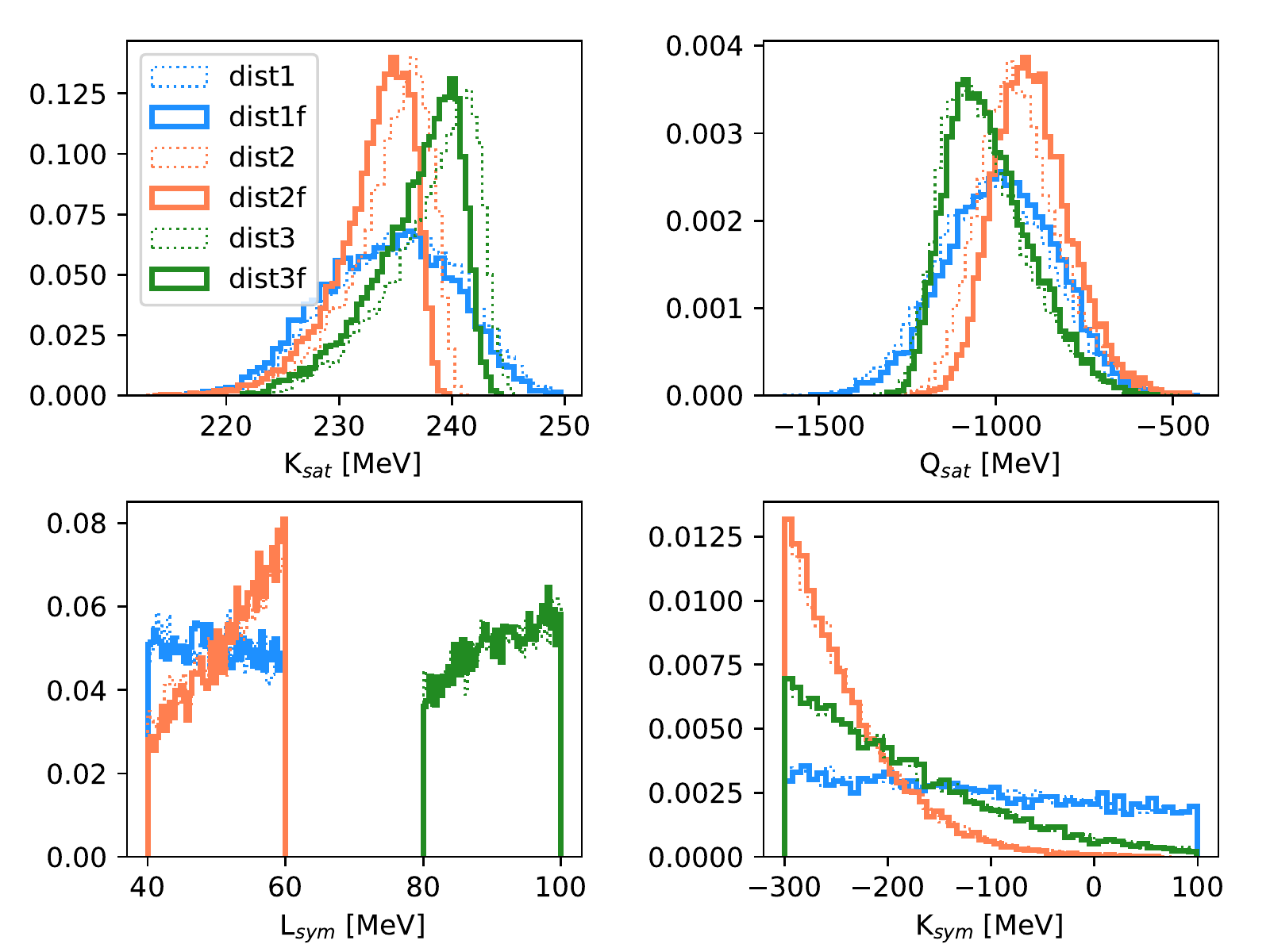}
\caption{One parameter marginalized distributions for the NEP parameters $K_\sat$, $Q_\sat$, $L_\sym$ and $K_\sym$. The distributions dist$i$f are shown in solid lines with same colors as in Fig.~\ref{fig:mcmc_corner}. They are compared to the distributions dist$i$ (without fine tuning to experimental nuclear masses) in thin dotted lines. The differences between dist$i$f and dist$i$ are small.
}
\label{fig:mcmc_NEP}
\end{figure}

The marginalized distributions for the NEP parameters are shown in Figs.~\ref{fig:mcmc_corner} and \ref{fig:mcmc_NEP}. The corner plot representation in Fig.~\ref{fig:mcmc_corner} shows the one parameter distributions on the diagonal and the correlation between the parameters off the diagonal, while in Fig.~\ref{fig:mcmc_NEP} we show a zoom of the one parameter distributions. We compare in the distributions obtained without the fictitious data for $^{132}$Sn (dist1f, blue) and with this fictitious data (dist2f, red). We also show the marginalized distribution when the slope of the symmetry energy $L_\sym$ is taken to be large and around 90~MeV (dist3f, green), as suggested by the analysis of PREX2 experimental data~\cite{Brendan2021}. The Gaussian distributions in dashed lines represent the expected distributions for these parameters from Ref.~\cite{Margueron2018a}. They are also given in the first raw of Tab.~\ref{tab:NEP}. 

Let us first remark that the value for the parameter $Q_\sat$ is very different from the expected values given in Tab.~\ref{tab:NEP}. The distribution for $Q_\sat$ is very similar for the three cases: it is peaked at around $\approx -950$~MeV with an uncertainty of about $150$-$200$~MeV. The value extracted from an analysis of models predictions, since there are no direct extraction from experimental data of this parameter, is expected to be quite different: $\approx 300\pm 400$~MeV~Ref.~\cite{Margueron2018a}. These values are extracted from an analysis over existing non-relativistic and relativistic phenomenological approaches. However, it was already noticed that the value of this parameter changes a lot from a type of nuclear interaction to another: about -350~MeV in average for Skyrme models, around 0 for relativistic mean-field (RMF) models and around 390~MeV for relativistic Hartree-Fock (RHF) ones. There is therefore a large model dependence of $Q_\sat$, which may be related to its correlation with $K_\sat$ as suggested in Ref.~\cite{Khan2013b}. The value preferred by the GMR data points toward a region which is orthogonal to any value of existing models. We can then deduce that in order to reproduce correctly several isotopic chains from Zr to Pb, including Sn isotopes, the required value for $Q_\sat$ is quite different from the typical values given in phenomenological approaches. So the possible origin of the difficulties faced by the usual phenomenological models in reproducing both the Sn and Pb isotopes could take its origin in the values of the NEP $Q_\sat$ in these models. To reproduce better Sn and Pb isotopes, more flexibility shall be given to these models, in particular the breaking of the correlation between $K_\sat$ and $Q_\sat$. For Skyrme models, this could come with an additional density-dependent term, or the '$t_3$' kind, as suggested in Ref.~\cite{Lesinski06}. 

The second remark is about the role of a fictitious measurement of the GMR energy in $^{132}$Sn. For simplicity, we assumed an accurate measurement as $E_\gmr(^{132}$Sn$)=14.8$~MeV, see dist2f. An uncertainty in $E_\gmr(^{132}$Sn$)$ will produce a result between the one suggested by dist1f and dist2f, except if the measurement is lower than the value we considered. Let us simplify the discussion of this fictitious data by not considering such a case. The role of this fictitious data for $E_\gmr(^{132}$Sn$)$ can be seen from the difference between dist1f (blue) and dist2f (red) distributions. While the isoscalar NEP are weakly impacted, the isovector NEP $K_\sym$ is largely impacted by the fictitious data: such a new measurement would shift the expected value for $K_\sym$ towards large and negative values. 

Note also that the value of $L_\sym$ is not constrained by the considered experimental values: $L_\sym$ fully explores the flat prior without specific structure and it is also not correlated to other NEP. There are however correlations between $K_\sat$ and $Q_\sat$, as well as between $K_\sat$ and $K_\sym$ and $Q_\sat$ and $K_\sym$. The distribution for $K\sat$ is more peaked than the empirical expectation (with a width of $\pm 20$~MeV). One of the reason is that there is correlation between the parameter $a_\surf$, controlling the density dependence of the surface energy, and $K_\sat$. In the present study we have fixed $a_\surf=-20$, as resulting from the typical value we obtained in the previous subsection. This parameter is however not fixed by any experimental data and including its uncertainty may contribute to widen the $K_\sat$ distribution. Another reason comes from the better agreement of our model with the experimental data, in comparison to other phenomenological approaches, e.g., Skyrme or RMF~\cite{Khan2013}. Since in our model we can fix the value of $Q_\sat$ independently of $K_\sat$, it results in a better description of the experimental $K_A$ values and the parameters $K_\sat$ and $Q_\sat$ are better determined, see Figs.~\ref{fig:mcmc_corner} and \ref{fig:mcmc_NEP}. In other words, the uncertainties in $Q_\sat$ impacts the one in $K_\sat$, as suggested in Ref.~\cite{Margueron2019}. Since $Q_\sat$ is better known from the present approach, it results that $K_\sat$ is also determined with a better accuracy.

\begin{figure}[t]
\centering
\includegraphics[scale=0.4]{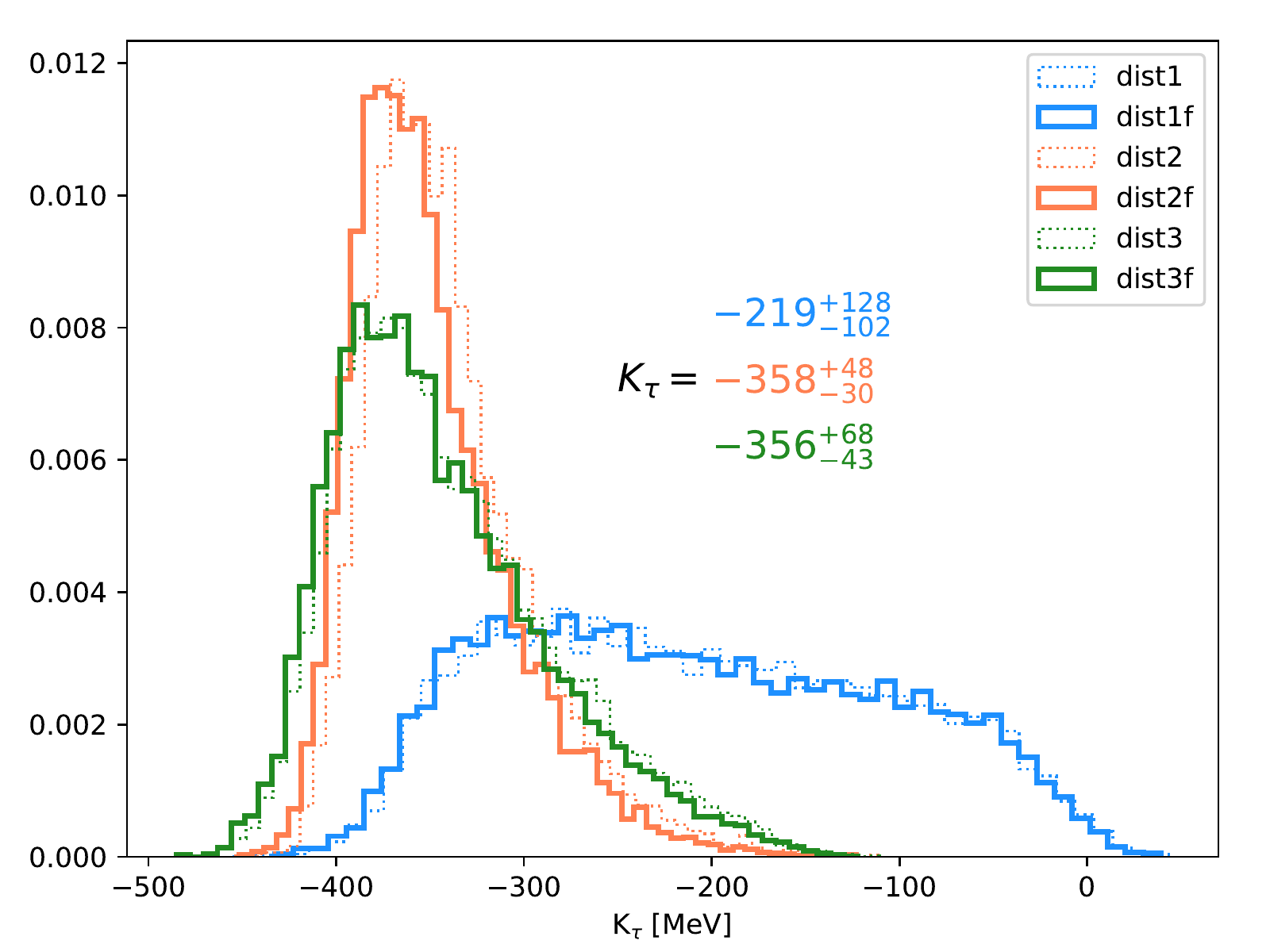}
\caption{Marginalized distributions for the parameter $K_\tau$. Same legend as in Figs.~\ref{fig:mcmc_corner} and \ref{fig:mcmc_NEP}. The centroids for $K_\tau$ are given in the figure for the cases dist$i$f.}
\label{fig:mcmc_Ktau}
\end{figure}

We represent in Fig.~\ref{fig:mcmc_Ktau} the marginalized distribution for the parameter $K_\tau$, defined from Eq.~\eqref{eq:ktau} for the cases dist$i$ (think dotted lines) and dist$i$f (thick solid lines). Without the fictitious GMR energy in $^{132}$Sn (dist1 and dist1f) the $K_\tau$ distribution is quite flat, while when the $^{132}$Sn fictitious data is considered, the $K_\tau$ distribution is better localized. For the value we considered including an accurate experimental data, we obtain $K_\tau\approx -358\pm 40$~MeV ($K_\tau\approx -356\pm 50$~MeV) for $L_\sym\approx 50\pm10$~MeV ($L_\sym\approx 90\pm10$~MeV). Here also, we note the relative independence of the $K_\tau$ distribution in the parameter $L_\sym$.

Our results also differ from others if we do not consider the fictitious data in $^{132}$Sn. The value $K_\tau\approx -550\pm 100$~MeV was extracted from the analysis of the Sn isotopic chain only (from $^{112}$Sn to $^{124}$Sn~\cite{Li2010}). Note also the value $K_\tau\approx-500\pm 50$~MeV extracted from the same experimental data, using different Skyrme Hamiltonians and RMF Lagrangians~\cite{Sagawa2007}. If we apply our analysis to the same data points as in Ref.~\cite{Li2010,Sagawa2007}, then we obtain  $K_\tau\approx-330\pm 120$~MeV and $K_\tau\approx-270\pm 100$~MeV if we impose to reproduce $K_A$ in Pb as well. Note however that when we consider the fictitious data in $^{132}$Sn, the value for $K_\tau$ become more peaked. This illustrates the role of isotopes with large isospin asymmetry in the determination of $K_\tau$. However, for these data on exotic nuclei to be effective, they need to be as accurate as the data obtained for stable nuclei.

\begin{figure}[t]
\centering
\includegraphics[scale=0.5]{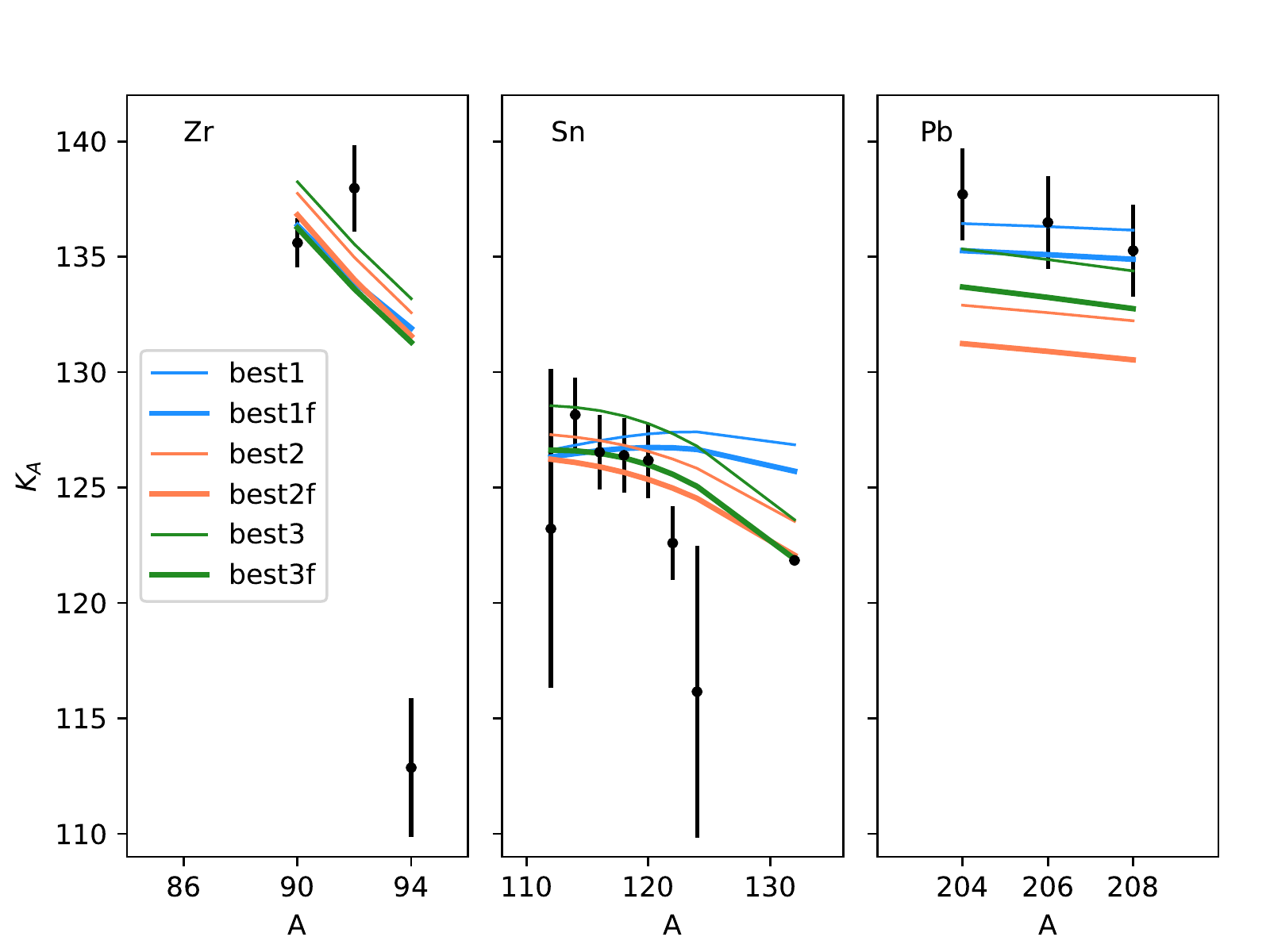}
\caption{Comparison of the best parameter set against the experimental data for Zr, Sn and Pb isotopes.}
\label{fig:mcmc_KA}
\end{figure}

Finally, we show in Fig.~\ref{fig:mcmc_KA} the comparison between the experimental values for $K_A$ and the values obtained with our best parameter set for each cases dist$i$ (thin lines) and dist$i$f (thick lines). Our eCLDM model is able to well reproduce the experimental points in Zr, Sn and Pb isotopes with a very good accuracy. Once again, this is possibly due to the large negative Q$_{sat}$ value, which points to a hint for solving the so-called Sn softness puzzle. Note that the experimental point in $^{94}$Zr is out of reach from our model. The difference between $K_A$ in $^{92}$Zr and $^{94}$Zr is too large to be reproduced. For this reason, we decided not to include $^{94}$Zr in our fit. We also advocate for a new measurement in $^{94}$Zr, since the present data is surprising.

In the case dist1 and dist1f, the best parameter sets provide a consistent description of the experimental value in Zr, Sn and Pb isotopes. Note however that the data in $^{124}$Sn is not very constraining in our case, since the uncertainty is large. Therefore, the evolution of $K_A$ over the Sn isotopic chain is quite flat in our model. The effect of including the fictitious data in $^{132}$Sn with small uncertainty, forces our model to decrease $K_A$ as function of $A$ in Sn isotopes (see dist2 and dist2f). The description of Pb isotopes, while still good, is slightly deteriorated. It is however restored with dist3 and dist3f, where a larger value for $L_\sym\approx 90\pm10$~MeV is explored. However, these results are still exploratory and no conclusion could be given without an accurate measurement of the GMR energy in $^{132}$Sn.

\section{Sound speed in nuclei and uniform matter}
\label{sec:sound_speed}

\begin{figure}
\centering
\includegraphics[width=0.49\textwidth]{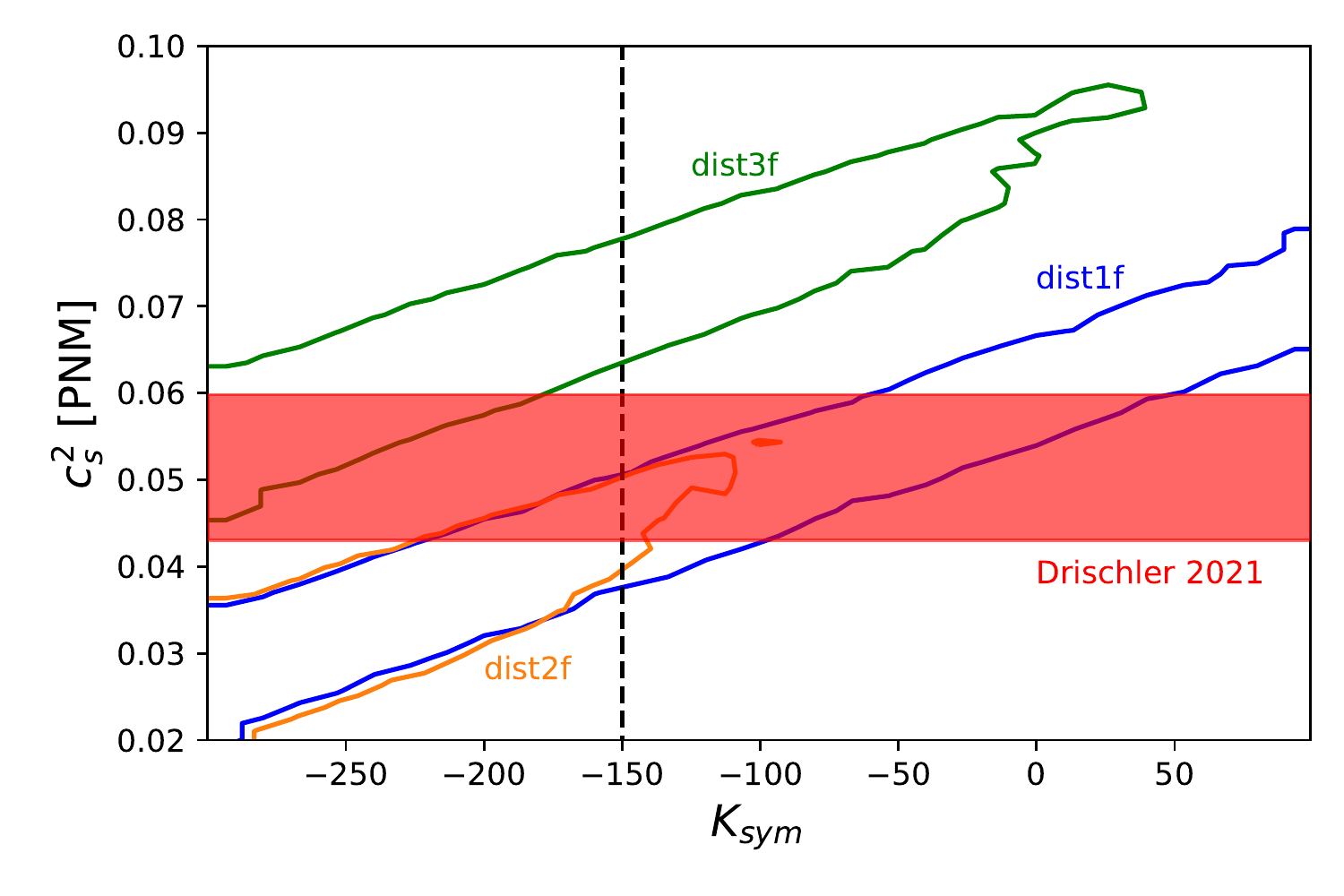}
\caption{Correlation of the sound speed in NM at $n_\sat$ with the parameter $K_{sym}$ for the three cases: dist1f, dist2f and dist3f. The horizontal red band (Drischler 2021) depicts the constraint on the sound speed from $\chi$EFT Hamiltonians calculated at $n_\sat$~\cite{Drischler:2020hwi}. The vertical line at $K_\sym=-150$~MeV separates the excluded values obtained from our Eq.~\eqref{eq:ksymvalue} (left side) from the authorized ones (right side).}
\label{fig:corr_c2_Ksym_exp}
\end{figure}

The sound speed is an important property in transport models~\cite{Danielewicz:2002pu}. It is interesting to address the effects of the nuclear properties, e.g., NEPs, on the sound speed in uniform matter and in finite nuclei. Moreover, the sound speed is an important ingredient in the calculation of the tidal deformation in binary neutron stars~\cite{Hinderer:2007mb,Flanagan:2007ix,Somasundaram:2021clp,Han:2018mtj}. Therefore a connection between the sound speeds in finite nuclei and infinite matter could help in constraining NS observables from nuclear experiments. 

The sound speed $c_s$ in a nuclear fluid is largely determined from the nuclear incompressibility in asymmetric matter. It is defined as~\cite{Blaizot80}
\begin{equation}
c_s^2 = \frac{K(n,\delta)}{ 9h(n,\delta)} \, ,
\end{equation}
where $h$ is the enthalpy per  particle $h=mc^2+e+P/n$. In uniform matter, we determine the sound speed from the following quantities $e_\um$~\eqref{eq:eAMEP}, $P_\um$~\eqref{eq:pum} and $K_\um$~\eqref{eq:kea}, while in finite nuclei, we use $e_A$~\eqref{eq:energy_nucleus}, $P_A$~\eqref{eq:pressure_nucleus} and $K_A$~\eqref{eq:KA}. All these quantities have been defined in previous sections and in the appendix A.

We show in Fig.~\ref{fig:corr_c2_Ksym_exp} the correlation between the speed-of-sound in NM at $n_\sat$ and the NEP $K_{\sym}$. We have used the posterior distributions corresponding to the three cases: dist1f, dist2f and dist3f for all the NEP (here $K_\sat$, $Q_\sat$, $K_{\sym}$ and $L_\sym$). The findings of Fig.~\ref{fig:corr_c2_Ksym_exp} suggest that a tight constraint on the value of the sound-speed in NM at around saturation density, could turn into a constraint of the value of $K_{\sym}$. With the advent of ab-initio calculations such as  $\chi$EFT~\cite{Drischler:2020hwi} it is possible to determine a band for the sound-speed in NM. In Fig.~\ref{fig:corr_c2_Ksym_exp}, the red band shows chiral EFT calculations for the sound speed in NM at $n_\sat$ obtained in Ref.~\cite{Drischler:2020hwi}. At $n_\sat$, the intersection of the red band ($\chi$EFT) and blue contour (dist1f) suggests that $-200\lesssim K_\sym\lesssim  50$~MeV. We have performed a similar analysis at $2n_\sat$ but it does not bring any additional information on $K_{\sym}$. 

It should be noted, from Fig.~\ref{fig:corr_c2_Ksym_exp}, that the inclusion of a fictitious data in $^{132}$Sn (see the contour Dist2f) may contribute to the reduction of the band width for the sound speed in NM, reducing the values for $K_\sym$ to be $K_\sym \lesssim -100$~MeV and $c_s^2\lesssim 0.055 c^2$ (instead of $0.06 c^2$). The case of dist3f is even more interesting: if only large values for $L_\sym\sim 90\pm10$~MeV compatible with PREX2 are authorized then the overlap between dist3f and $\chi$EFT occurs in the forbidden region for $K_\sym$. In other word, there is no overlap between dist3f and $\chi$EFT. The sound speed in NM therefore contributes to exclude large values for $L_\sym$, as suggested by PREX2.

\begin{figure}
\centering
\includegraphics[width=0.49\textwidth]{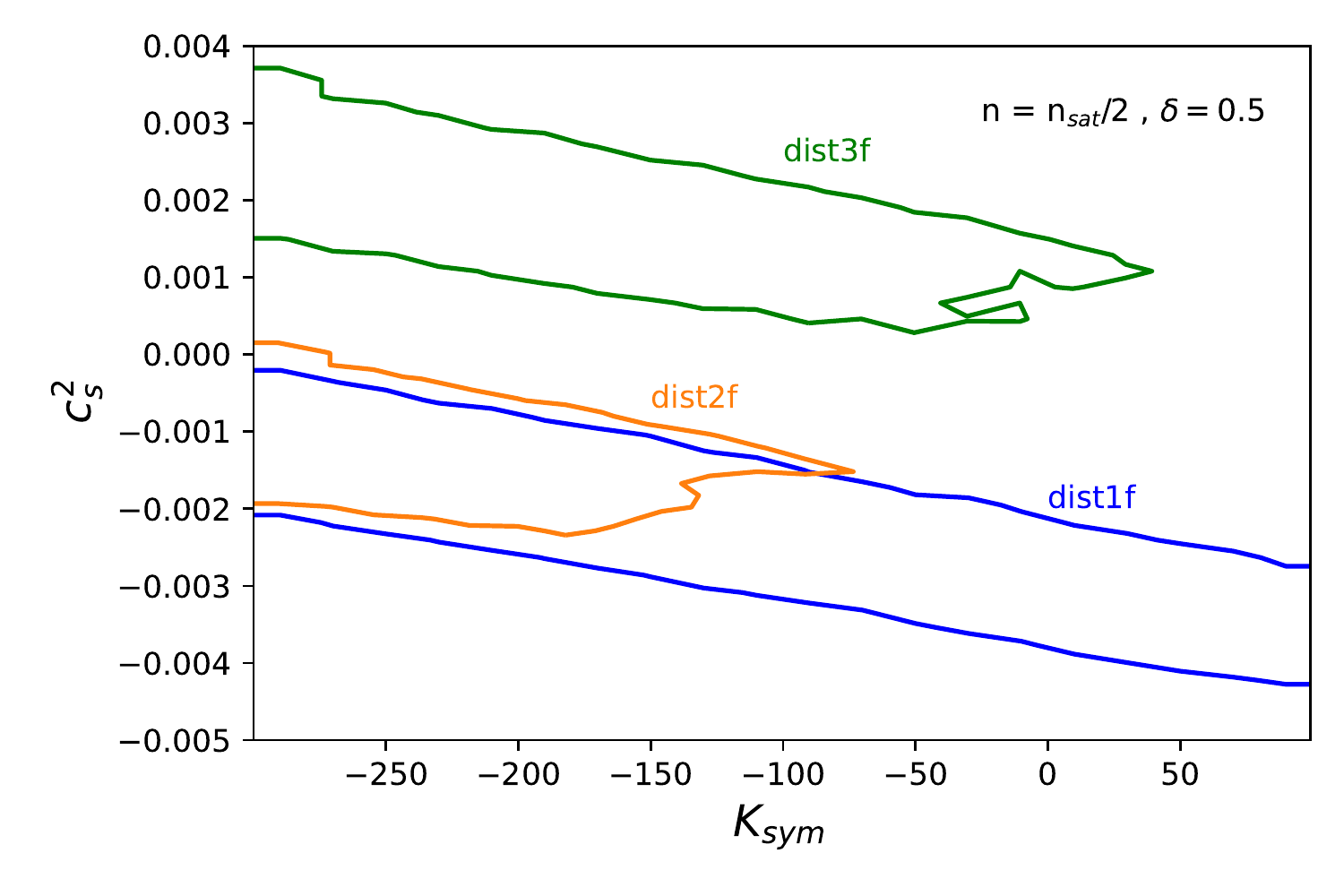}
\caption{Similar to Fig.~\ref{fig:corr_c2_Ksym_exp} but at half saturation density and with $\delta=0.5$}
\label{fig:corr_c2_Ksym_exp_asym}
\end{figure}

It is also relevant to explore the correlation between the sound-speed and $K_{\sym}$ in cases similar to what exists in heavy ion collisions at the Fermi energy. In Fig.~\ref{fig:corr_c2_Ksym_exp_asym}, we fix the density to be $n_\sat/2$ and isospin asymmetry parameter $\delta=0.5$. Interestingly we see that, in contrast with the NM case ($\delta=1$) shown in Fig.~\ref{fig:corr_c2_Ksym_exp}, the correlation is negative (anti-correlation). This is due to the dominant contribution of the pressure to the enthalpy, for which $K_\sym$ contributes to a large extent: $K_\sym$ contributes to the first power in the density parameter $x$ to the pressure, while only to the second power to the energy per particle. The isoscalar contribution to the pressure is small in the vicinity of saturation density. Since the leading order impact of $K_\sym$ is an odd power in $x$, it has an opposite correlation below saturation density, as compared to above. 
As in Fig.~\ref{fig:corr_c2_Ksym_exp}, we see that the uncertainty in $L_\sym$ plays a large role, as can be inferred by comparing dist3f with the other cases. The uncertainty induced by $L_\sym$ is of similar magnitude as the one originating from $K_\sym$.
We can thus conclude that tighter constraints on both $L_{\sym}$ and $K_\sym$ will reduce the uncertainty in the sound speed.

\begin{figure}
\centering
\includegraphics[width=0.45\textwidth]{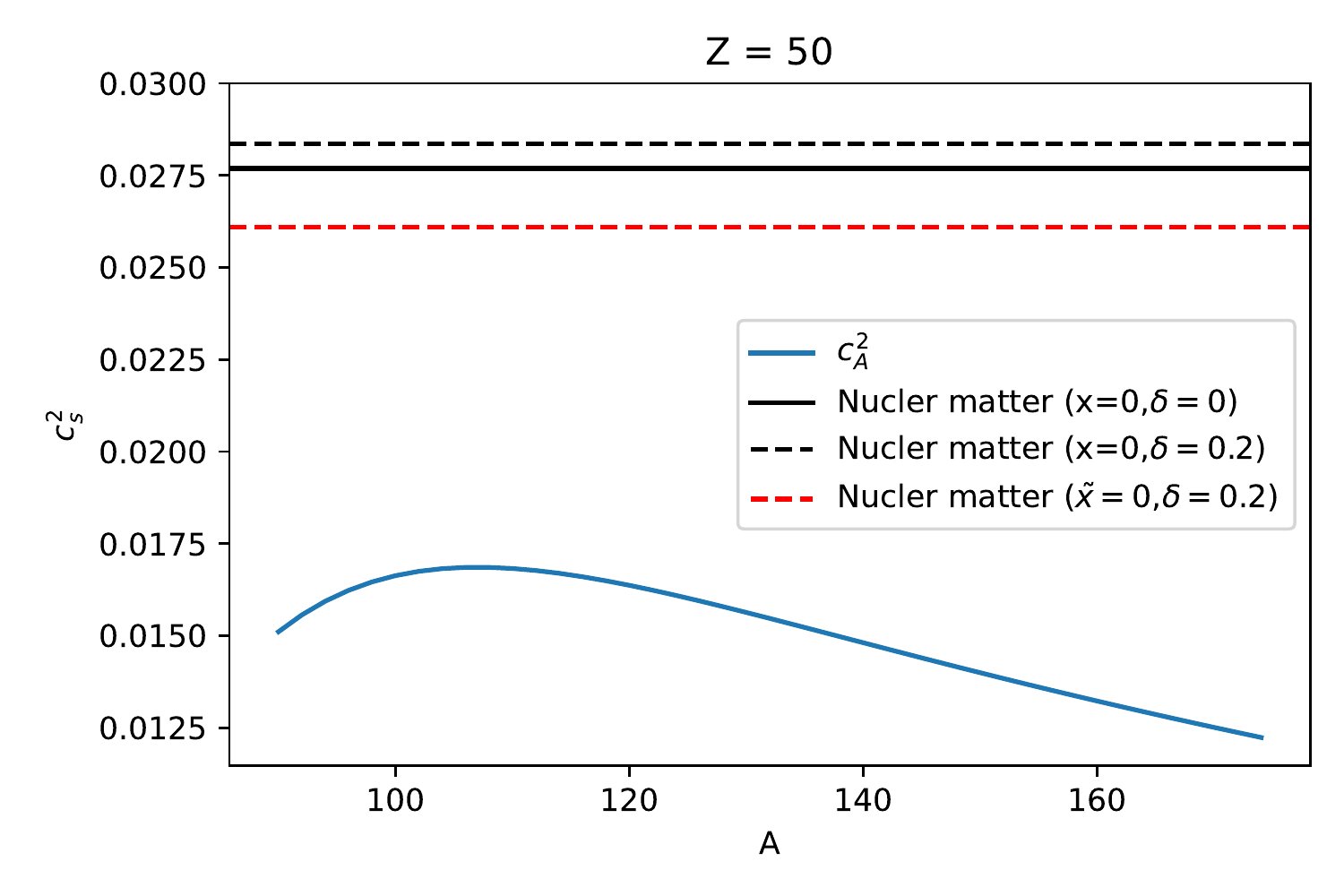}
\includegraphics[width=0.49\textwidth]{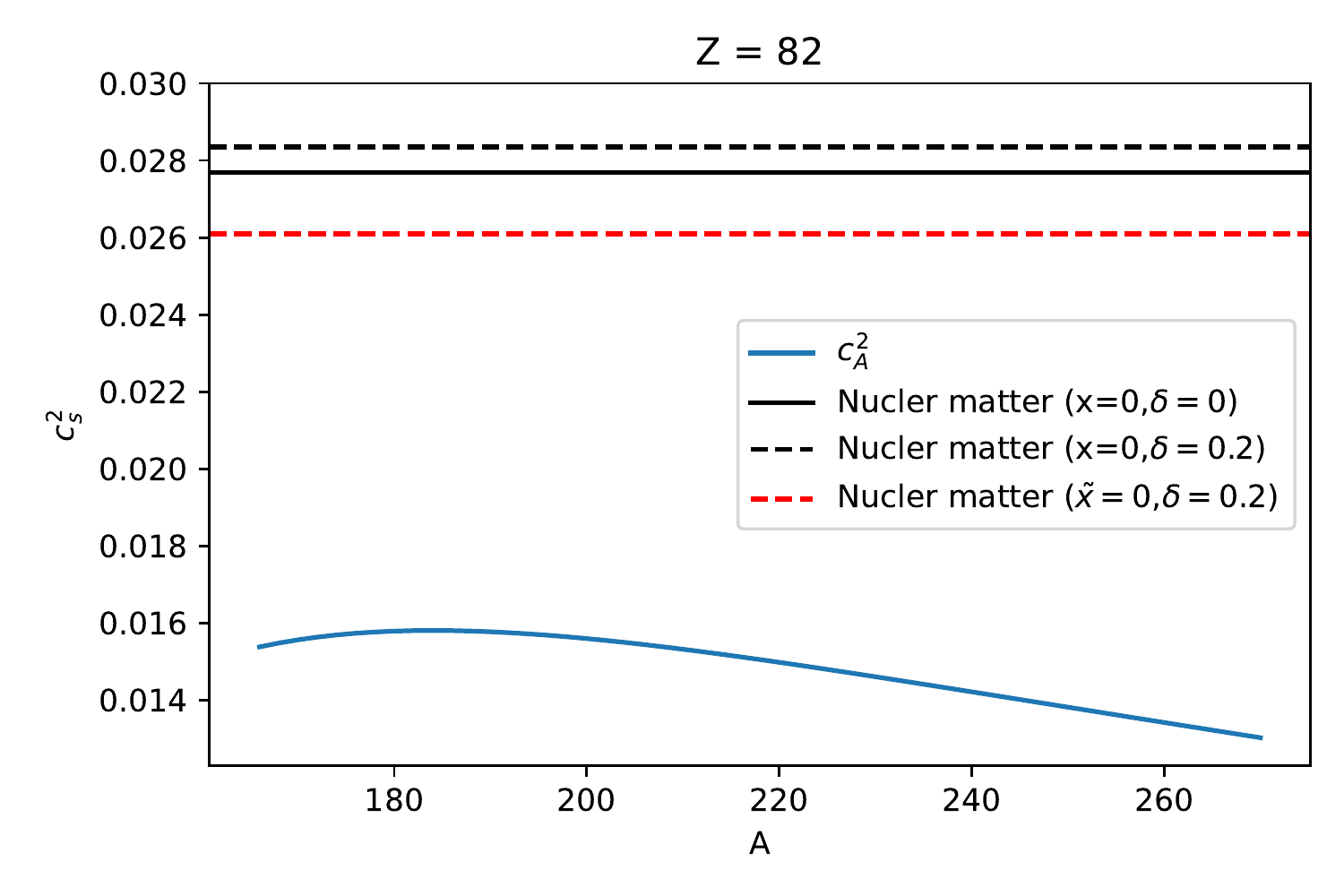}
\caption{The speed of sound in finite nuclei is shown (blue curve) as a function of $A$ for Sn (top) and Pb (bottom). The black lines represent nuclear matter results as indicated in the legend.
}
\label{fig:c2_A_Z50}
\end{figure}

We now come back to finite nuclei, where FS terms also play a role in sound speed. These FS terms impact the connection between the sound-speed in finite nuclei and the sound-speed in nuclear matter. In Fig.~\ref{fig:c2_A_Z50}, we show the sound-speed in finite nuclei as a function of $A$ for two isotopic chains: Sn and Pb. For this calculation, we have used the SLy5 interaction. In both panels, the sound speed in infinite matter at $n_\sat$ is shown as black horizontal lines. The solid black line represents SM. The red dashed line represents AM with $\delta=0.2$ and for $n_\eq=0.157$~fm$^{-3}$. So the effect of asymmetry itself is to slightly increase the sound speed, while shifting down the equilibrium density from SM to AM reduces the sound speed. The main differences between the blue curve and the straight lines, originate from the contribution of the FS terms in finite nuclei. There is a factor approximately 2 between uniform matter and finite nuclei. The same difference has been observed between $K_{\sat}$ and $K_A$, see for instance the middle panel of Fig.~\ref{fig:satEP}. Interestingly, we see that the deviation between the black and the blue lines increases with $A$ due to the fact that nuclei get more and more neutron rich, and therefore the cluster density decreases. At much larger $A$ (above the values shown in the figures), the FS terms finally decrease in size and at the limit $A\rightarrow\infty$ finite nuclei and uniform matter results do get closer.

\section{Conclusions}

In this work we have explored various ways to encode the density and isospin asymmetry dependence of the incompressibility in nuclear matter. We have discussed the dominant contribution of $L_\sym$ in the determination of the isospin dependence of the incompressibility modulus. A better knowledge of the incompressibility modulus in AM requires therefore an accurate knowledge of $L_\sym$. In finite nuclei, by introducing an extended CLDM (eCLDM) adding a density dependence to the surface tension proportional to $x_\cl^2$, where $x_\cl = (n_\cl -n_\sat)/3n_\sat$, we were able to provide a unified macroscopic model for nuclear masses and incompressibility modulus. We have then rederived $K_A$ from the eCLDM framework along the lines originally suggested by Blaizot~\cite{Blaizot80}. In this way, the contribution of the new density-dependent term, in the surface tension to $K_A$, is explicitly shown in the equations.

We have compared the predictions of the eCLDM for the nucleus incompressibility $K_A$ with microscopic calculations and experimental data. Thanks to the flexibility of the meta-model, a sensitivity analysis on the impact of individual nuclear empirical parameter is made.  As expected, the isoscalar channel influences the absolute values of the energies while the isoscalar one impact the slope of the $K_{A}$ as function of the isospin asymmetry. A full exploration in the parameter space formed by $K_\sat$, $Q_\sat$, $L_\sym$ and $K_\sym$ is also performed, showing that the parameter $Q_\sat$ must be approximately $Q_\sat\approx 950\pm 200$~MeV to reconcile the experimental GMR energies measured in Zr, SN and Pb isotopes. Since this suggested value is different from the ones of phenomenological forces, we then suggest a possible explanation of the origin of the difficulties these forces faces in reproducing the experimental data on $K_A$, on both Sn and Pb nuclei. 

In addition, we explore the impact of a fictitious accurate measurement for the GMR energy in $^{132}$Sn. We show that with such a measurement, the value of $K_\sym$ and $K_\tau$ would be much better determined than they are with the present data.

We have also derived two new constraints on $K_\sym$:
\begin{itemize}
\item From the equilibrium density: $K_\sym\geq-K_\sat+\sqrt{2L_\sym Q(\delta=1)}\sim-150$~MeV and $Q_\sym>-Q_\sat$.
\item From the confrontation of our prediction for the sound speed in NM with the $\chi$EFT, we found $-200\lesssim K_\sym \lesssim 50$~MeV. This constraint could be more accurate if a measurement of the GMR energy in $^{132}$Sn is known.
\end{itemize}
Let us remark that the constraint on $Q_\sym$ combined with the MCMC exploration for $Q_\sat$ leads to the following consequence: $Q_\sym \gtrsim 950\pm 200$~MeV.

In conclusion, the present work suggests a new way to analyze the experimental $K_A$ and to extract the values of the NEP $K_\sat$, $Q_\sat$, $L_\sym$ and $K_\sym$, which are the most influential ones. This method is comparable to the microscopic Hartree-Fock one, except that it does not describe shell effects. These shell effects are however reduced by the treatment of the pairing, as shown in the microscopic Hartree-Fock Bogoliubov calculations \cite{GARG2018}. The advantage of our method is that we use the flexible nuclear meta-model to simulate the role of the nuclear interaction. At variance with phenomenological forces, the nuclear meta-model is able to freely choose the best NEP which describe the experimental data. We found that the data favors a large and negative value for $Q_\sat$ which is not possible with phenomenological forces. We then suggest a possible origin for the observed limitations of these forces.

We note that our ability to extract information on $Q_\sat$ and $K_\sym$ from finite nuclei, is based on the variation in density and isospin asymmetry explored by the isotopes defining the loss function $\chi_E$. For simplicity we have based our analysis on the results of an eCLDM where the densities and isospin asymmetries are taken flat in finite nuclei. This is clearly an important feature which has to be improved in the future. One may think for instance in implementing the meta-model in a modeling of finite nuclei with better density profiles compared to the eCLDM. Further works in this direction are therefore envisioned.

\acknowledgements

J.M. and R.S. are supported by CNRS grant PICS-08294 VIPER (Nuclear Physics for Violent Phenomena in the Universe) and the CNRS IEA-303083 BEOS project. All authors are grateful to the CNRS/IN2P3 NewMAC master-project, and benefit from PHAROS COST Action MP16214. This work is supported by the LABEX Lyon Institute of Origins (ANR-10-LABX-0066) of the \textsl{Universit\'e de Lyon} for its financial support within the program \textsl{Investissements d'Avenir} (ANR-11-IDEX-0007) of the French government operated by the National Research Agency (ANR). GG is supported by Fonds de la Recherche Scientifique (F.R.S.-FNRS) and the Fonds Wetenschappelijk Onderzoek - Vlaanderen (FWO) under the Excellence of Science (EOS) Projects nr O022818F and O000422F.

\appendix


\section{Derivation of the incompressibility $K_A$ in finite nuclei}
\label{app:KAderivation}

In this appendix we derive $K_A$ in finite nuclei and obtain an expression similar to the one obtained by Blaizot~\cite{Blaizot80}, but for the eCLDM approach and where we have introduced the NEP explicitly. We detail the derivation step by step, starting with the definition for the nuclear binding energy, going to the pressure in the nucleus to have in the end a clear expression for the incompressibility. In the last section of this appendix we take advantage of the present approach to analyse the impact of the NEP to reproduce $K_A$.

\subsection{Energy per particle: $e_A$}

We define the binding energy $e_A$ for the nucleus $A$ in the CLDM as, 
\begin{equation}
e_{A} \equiv  e_{A,\um} + e_{A,\FS}
\label{eq:energy_nucleus}
\end{equation}
where the uniform matter energy $e_{A,\um}$ is defined from the symmetric matter and symmetry energy terms $e_{\sm}$~\eqref{eq:eSMEP} and $e_{\sym}$~\eqref{eq:eSymEP} as $e_{A,\um}(n_\cl,\delta_\cl) \equiv e_{\sm}(n_\cl) + e_{\sym}(n_\cl)\delta_\cl^2$, where $n_\cl$ and $\delta_\cl$ are the equilibrium density and isospin asymmetry $\delta_\cl=(N-Z)/A$ of a given nucleus.
The density $n_\cl$ is obtained assuming that the nucleus is at mechanical equilibrium, {\it i.e.} $P_{A} = 0$, see discussion in the next subsection.

The finite size contribution is defined as, 
\begin{eqnarray}
e_{A,\FS} &\equiv& e_{A,\surf} + e_{A,\coul}
\label{eq:efs}
\end{eqnarray}
where we consider only the surface and Coulomb terms in the present work. The contributions originating from higher order terms in the leptodermous expansion are disregarded in this analysis, where we present an eCLDM with a density-dependent surface energy. However, they shall be studied in a future work. Note also that by considering only the FS terms as given in Eq.~\eqref{eq:efs}, our equations are consistent with the seminal paper by Blaizot~\cite{Blaizot80} (see for example Eq.~(2.17)). 

Considering the direct Coulomb contribution only, as well as a uniform charge distribution in the nucleus, the Coulomb energy reads, 
\begin{eqnarray}
e_{A,\coul} &=& \mathcal{C}_\coul \frac{3}{5}\frac{Z^2e^2}{R_A} \frac{1}{A}, 
\end{eqnarray}
where the nucleus radius is  $R_A = r_\cl A^{1/3}$ and $r_\cl = (3 / 4\pi n_\cl)^{1/3}$. The parameter $\mathcal{C}_\coul$, which is fitted on experimental nuclear masses (see Ref.~\cite{Grams2022a} for details on the fit procedure), represents an effective way to incorporate the effect of exchange as well as of the surface, on the Coulomb energy. In the present fit, the experimental masses are corrected by the odd-even mass staggering as $\tilde{E}_{\rm ex}^i = E_{\rm ex}^i - \Delta E_{\rm ex}^i$, with
\begin{equation}
\Delta E_{\rm ex}^i = \left[\Delta_\sat + \Delta_\sym \left(\frac{N_i-Z_i}{A_i}\right)^2 \right] \, A_i^{-1/3} \, \delta(N,Z) \, .
\end{equation}
where $\delta(N,Z) = 1$ if $N$ and $Z$ are odd, $0$ if either $N$ or $Z$ is odd, and $-1$ if both $N$ and $Z$ are even~\cite{BohrMottelson1969}. The parameters $\Delta_\sat$ and $\Delta_\sym$ are varied together with the CLDM parameters $\mathcal{C}_i$ in the fit to the experimental masses. We show the optimal CLDM parameters $\mathcal{C}_i$ and the odd-even mass staggering parameters for each Skyrme model in Table~\ref{tab:FS}. Note that the values we obtain for $\Delta_\sat$ and $\Delta_\sym$ are similar to the ones determined in Ref.~\cite{Vogel1984}.

\begin{table*}[t]
\centering
\tabcolsep=0.1cm
\def\arraystretch{1.5}
\begin{tabular}{cccccccccc}
\hline\hline
Model                & BSK14 & BSK16 & F0 & LNS5 & RATP & SGII & SKI2 & SKO & SLy5 \\
\hline
$\mathcal{C}_\coul$       & 0.93/0.94 & 0.95/0.95 & 0.93/0.94 & 0.91/0.91 & 0.95/0.95 & 0.92/0.92 & 0.93/0.93 & 0.93/0.93 & 0.94/0.94 \\
$\mathcal{C}_\surf$       & 1.03/1.02 & 1.07/1.06 & 1.09/1.08 & 0.98/0.97 & 1.07/1.06 & 0.97/0.96 & 1.00/1.00 & 1.03/1.03 & 1.07/1.06 \\
$\mathcal{C}_{\surf,\sym}$& 0.98/0.94 & 0.92/0.87 & 1.30/1.24 & 0.93/0.90 & 0.81/0.76 & 0.58/0.54 & 1.40/1.45 & 1.26/1.25 & 1.31/1.25 \\
$\Delta_{\sat}$ (MeV)   & 12.5/12.4 & 12.1/12.0 & 12.5/12.4 & 12.8/12.7 & 11.9/11.8 & 12.1/12.0 & 13.3/13.4 & 13.0/12.9 & 12.2/12.1 \\
$\Delta_{\sym}$ (MeV)   & -37.5/-34.6 & -22.1/-19.5 & -38.3/-34.8 & -51.9/-49.8 & -14.8/-12.4 & -24.9/-21.7 & -73.2/-77.1 & -58.0/-57.2 & -42.4/-38.9 \\
$\sqrt{\chi^2}$ (MeV)      & 3.3/3.2 & 3.3/3.1 & 3.4/3.3 & 3.6/3.4 & 3.3/3.1 & 3.5/3.3 & 3.7/3.5 & 3.6/3.3 & 3.4/3.3 \\
\hline\hline
\end{tabular}
\caption{Optimized finite size parameters and loss function $\sqrt{\chi^2}$ with eCLDM/CLDM. For eCLDM we use $a_\surf = -20.0$.}
\label{tab:FS}
\end{table*}

The surface energy is given by, 
\begin{eqnarray}
e_{A,\surf} &=&  \mathcal{C}_\surf \, 4 \pi \sigma_\surf R^2_A \frac{1}{A}.
\end{eqnarray}

\begin{table}[t]
\centering
\tabcolsep=0.8cm
\def\arraystretch{1.5}
\begin{tabular}{ccc}
\hline\hline
$\sigma_\mathrm{surf,sat}$ & $\sigma_\mathrm{surf,sym}$ & $p_\surf$ \\
MeV~fm$^{-2}$ & MeV~fm$^{-2}$ & \\
\hline
1.1 & 2.3 & 3.0 \\
\hline\hline
\end{tabular}
\caption{Standard surface parameters for the CLDM considered in this work. Note the associated value $b_\surf=29.9$ deduced from Eq.\eqref{eq:sigmasurf}. }
\label{table:stdParam}
\end{table}

In the CLDM approach the surface tension is usually approximated by the following formula~\cite{LattimerSwesty1991},
\begin{equation}
  \sigma_\surf(I_\cl)\approx  \sigma_{\surf, \sat} \frac{2^{p_\surf+1}+b_\surf}{Y_{p,cl}^{-p_\surf}+b_\surf+(1-Y_{p,cl})^{-p_\surf}}, 
  \label{eq:sigma_old}
\end{equation}
where $Y_{p,\cl} = Z_\cl/A_\cl = (1-I_\cl)/2$, $I_\cl = (N_\cl - Z_\cl)/A_\cl $  and $\sigma_{\surf, \sat}$ is a parameter that determines the surface tension of symmetric nuclei. The isospin dependence is controlled by the parameters $b_\surf$ and $p_\surf$. Fixing the parameter $\sigma_{\surf, \sat}$ to an average value, see Tab.~\ref{table:stdParam}, the parameters $\mathcal{C}_\surf$ and $b_\surf$ are fitted from the nuclear chart, while the parameter $p_\surf$ is usually fixed to a value close to $\sim 3$~\cite{LattimerSwesty1991}, since it controls the isospin dependence of the surface energy for large asymmetries, which are not reached in finite nuclei. 

For small asymmetries we could expand $\sigma_\surf(I_\cl)$ as, 
\begin{equation}
\sigma_\surf(I_\cl)\approx \sigma_{\surf, \sat}-\sigma_{\surf,\sym} I_\cl^2
\label{eq:surfI}
\end{equation}
with
\begin{equation}
\sigma_{\surf,\sym} = \sigma_{\surf, \sat} \frac{2^{p_\surf} p_\surf(p_\surf+1)}{2^{p_\surf+1}+b_\surf}\, .
\label{eq:sigmasurf}
\end{equation}
Eq.~\eqref{eq:sigmasurf} relates the parameter $b_\surf$ to the surface symmetry energy $\sigma_{\surf, \sym}$, see also Ref.~\cite{Grams2022a} for more details. We fit the isoscalar and isovector surface parameters from the experimental nuclear masses. The standard surface parameters in the CLDM approach are given in Table \ref{table:stdParam}. The optimized parameters $\mathcal{C}_\coul$, $\mathcal{C}_\surf$ and $\mathcal{C}_{\surf,\sym}$ are given in Table \ref{tab:FS} for the different NEP used in the present work, together with the respective $\sqrt{\chi^2}$, where $\chi^2 = \frac{1}{N} \sum_{i=1}^{N} ( E_{\rm exp}^i - E_{\rm A}^i )^2  $.  $E^i_{\rm exp}$ are the experimental masses, $E^i_A$ are the predictions for the CLDM/eCLDM models for given nucleus $i$ and  $N=3375$ is the number of considered nuclei from the the 2020 Atomic Mass Evaluation (AME) \cite{AME2020}.

The novelty of the present work is the introduction of a density dependent surface tension, see Eq.~\eqref{eq:sigmaeCLDM}.
It should be noted that we have chosen the exponent of the density-dependent term in Eq.~\eqref{eq:sigmaeCLDM} to be two. The reason is twofold: first, it approximately satisfies the stationarity of the surface tension w.r.t the density, see Ref.~\cite{Blaizot80} for more details, and second, with such a power, it directly contributes to the incompressibility modulus in finite nuclei. Note that a correction proportional to $x_\cl$ has been suggested in Ref.~\cite{Iida04}, and analysed in view of its impact on the neutron skin. However, such a term does not satisfies the requested stationary of the surface tension and does not contribute to the incompressibility in finite nuclei.

\subsection{Pressure in finite nuclei: $P_A$}

The pressure $P_A$ in finite nuclei is defined as,
\begin{equation}
P_A \equiv n_\cl^2\frac{\partial e_A }{\partial n_\cl}\bat_A,
\label{eq:pressure_nucleus}
\end{equation}
which can be decomposed into a bulk term, originating from uniform matter and a finite size contribution: $P_A = P_{A,\um} + P_{A,\FS}$. The bulk term is decomposed into SM and isospin asymmetry contributions, as in Eq.~\eqref{eq:pum}: $P_{A,\um}=P_{A,\sm}+P_{A,\sym}\delta^2$, taking $P_{A,\sm}=P_{\sm}(n_\cl)$ and $P_{A,\sym}=P_{\sym}(n_\cl)$.

Note that the functions of $R_A$ in the binding energy also contribute to the pressure as,
\begin{equation} 
P_A = 
-\frac{R_A n_\cl}{3}\frac{\partial e_A}{\partial R_A}\Bigr|_{A} \, ,
\label{eq:PA:RA}
\end{equation}
since the partial derivative w.r.t. $n_\cl$ at fixed $A$ is equivalent to a partial derivative w.r.t. $R_A$, with appropriate factor, see appendix \ref{app:nclrcl_deriv}.

The finite size pressure is $P_{A,\FS} = P_{A,\surf} + P_{A,\coul}$, with the Coulomb pressure term derived as
\begin{equation}
P_{A,\coul} =  \frac{\mathcal{C}_\coul}{5}\frac{Z^2e^2 n_\cl}{R_A}\frac{1}{A} .
\end{equation}

The surface term is decomposed into two contributions 
\begin{equation}
P_{A,\surf} = P_{A,\surf}^\cldm + P_{A,\surf}^\dd 
\end{equation}
where the first term is the usual CLDM contribution, while the second term originates from the new density-dependent (DD) term. They are defined as,
\begin{eqnarray}
P_{A,\surf}^\cldm &=& - \frac{\mathcal{C}_\surf}{3} \, 8 \pi \sigma_\surf R_A^2 n_\cl\frac{1}{A} \, , \\
P_{A,\surf}^\dd &=&  \mathcal{C}_\surf \, 4 \pi R_A^2 n_\cl^2\frac{\partial \sigma_\surf}{\partial n_\cl}\frac{1}{A} \, .   
\end{eqnarray}
Note that since $\partial \sigma_\surf / \partial n_\cl \propto x_\cl\approx 0$, the contribution of the new DD term to the pressure is small. 

Numerically, the cluster density $n_\cl$ is obtained from the mechanical stability condition $P_A=0$, using the Newton-Raphson algorithm with $n_\eq^\um$ as starting solution.

\subsection{Incompressibility in finite nuclei: $K_A$}
\label{sec:KA}

The incompressibility $K_A$ in finite nuclei is defined as,
\begin{eqnarray} 
K_A &\equiv&  9 n_\cl\frac{\partial^2  \epsilon_{A}}{\partial n_{\cl}^2}\Bigr|_{A}, 
\label{eq:KA}
\end{eqnarray}
with the energy density given by $\epsilon_{A} = e_A n_\cl$. Similarly to the energy and the pressure, the linearity of the derivative operator allows to decompose the incompressibility $K_A$ in finite nuclei as bulk and FS terms,
\begin{eqnarray}
K_A &=& K_{A,\um} + K_{A,\FS}\, ,
\label{eq:KAUM_FS}
\end{eqnarray}
where $K_{A,\um}=K_{A,\sm}+K_{A,\sym}\delta^2$, $K_{A,\sm}=K_\sm(n_\cl)$ and $K_{A,\sym}=K_\sym(n_\cl)$.

The finite size contribution to the incompressibility are given as $K_{A,\FS} = K_{A,\surf} + K_{A,\coul}$, where the Coulomb term reads,
\begin{eqnarray}
K_{A,\coul} = \mathcal{C}_\coul  \frac{12}{5}\frac{Z^2e^2}{R_A}\frac 1 A\, ,
\label{eq:Kcoul}
\end{eqnarray}
and the surface term is expressed as
\begin{eqnarray}
K_{A,\surf} &=& \mathcal{C}_\surf \Big[ - 8\pi R_A^2\sigma_\surf + 24 \pi n_\cl R_A^2\frac{\partial \sigma_\surf}{\partial n_\cl} \nonumber \\ 
&& + 36 \pi n_\cl^2 R_A^2\frac{\partial^2 \sigma_\surf}{\partial n_\cl^2}\Big]\frac 1 A \, .
\label{eq:Ksurf}
\end{eqnarray}

\subsection{Re-expression of $K_A$}

The first time a CLDM model was used to compute the incompressibility of nuclei goes back to the seminal work of Blaizot~\cite{Blaizot80}.
In order to compare our expression for $K_A$ with his work, we dedicate this section to re-write our equations and obtain the equivalent of Eq. (6.3) of Ref.~\cite{Blaizot80}.

In finite nuclei, the density is different from the saturation density due to the contribution of FS and isospin asymmetry terms. If the density parameter $x_\cl$ remains small, Blaizot suggested to express $K_A$ as~\cite{Blaizot80},
\begin{equation}
K_{A} = K_{\sat} + \delta^2 \bar{K}_{A,\sym} + \bar{K}_{A,\FS}\, .
\label{eq:kanew}
\end{equation}

The new terms $\bar{K}_{A,\sym}$ and $\bar{K}_{A,\FS}$ incorporate, in addition to the contribution $K_{A,\sym}$ and $K_{A,FS}$, the shift in density between $n_\cl$ and $n_\sat$, see Appendix \ref{sec:xClExpression}, and more specifically Eq.~\eqref{eq:xClinPress}. To do so, we consider the expression for $K_{A,\sm}$ up to the linear order in $x_\cl$ from Eq.~\eqref{eq:KApowerx}, where the expression for $x_\cl$ in terms of $P_{A,\sym}$ and $P_{A,\FS}$ from Eq.~\eqref{eq:xClinPress} is injected:
\begin{eqnarray}
K_{A,\sm} &=& K_{\sat} - \frac{3 P_{A,\sym}}{ n_{\cl} K_\sat}\delta^2\left(12 K_\sat + Q_\sat \right) \nonumber \\
&&\hspace{1cm} - \frac{3 P_{A,\FS}}{ n_{\cl}K_\sat}\left(12 K_\sat + Q_\sat \right) \, .
\label{eq:newKbulkVol}
\end{eqnarray}
Re-ordering the different terms into $K_A$ gives Eq.~\eqref{eq:kanew} where
\begin{eqnarray}
\bar{K}_{A,\sym} &=& K_{A,\sym} - \frac{3 P_{A,\sym}}{ n_{\cl}K_\sat}\left(12 K_\sat + Q_\sat \right) \, ,\\
\bar{K}_{A,\FS} &=& K_{A,\FS} - \frac{3 P_{A,\FS}}{ n_{\cl}K_\sat}\left(12 K_\sat + Q_\sat \right)  \, .
 \end{eqnarray}
At order $o(x_\cl)$, the term $\bar{K}_{A,\sym}$ can be expressed as
\begin{equation}
\bar{K}_{A,\sym} \approx K_{\sym} - L_{\sym} \left(6 + \frac{Q_\sat}{K_\sat}\right) = K_\tau \, .
\end{equation}

The FS terms could be decomposed into the Coulomb and surface contributions. The Coulomb term reads,
\begin{equation}
\bar{K}_{A,\coul} = - \frac{3\mathcal{C}_\coul}{5}\frac{Z^2e^2}{R_A}\frac 1 A \left(8 +\frac{Q_\sat}{K_\sat}\right) = \bar{K}_{\coul}Z^2A^{-4/3} \, ,
\label{eq:Kcoulf}
\end{equation}
with
\begin{eqnarray}
\bar{K}_{\coul} &=& -  \frac{3\mathcal{C}_\coul}{5}\frac{e^2}{r_0} \left( 
8 +\frac{Q_{\sat}}{K_{\sat}} \right) \, .
\end{eqnarray}
The surface term reads,
\begin{eqnarray}
\bar{K}_{A,\surf} &=& \Big(\bar{K}_{\surf}^{\cldm}+\bar{K}_{\surf}^{\dd,\dot{\sigma}} + \bar{K}_{\surf}^{\dd,\ddot{\sigma}}\Big)A^{-1/3} \, .
\label{eq:KsurfTilda}
\end{eqnarray}
with
\begin{eqnarray}
\bar{K}_{\surf}^{\cldm} &=&  \mathcal{C}_\surf 8 \pi r^2_{\cl} \sigma_\surf\left(11 +  \frac{Q_{\sat}}{K_{\sat}} \right) \, , \\
\bar{K}_{\surf}^{\dd,\Dot{\sigma}} &=& - \mathcal{C}_\surf 12\pi n_\cl r_\cl^2 \frac{\partial \sigma_\surf }{\partial n_\cl}\left( 
10 +   \frac{Q_{\sat}}{K_{\sat}} \right)
\\
\bar{K}_{\surf}^{\dd,\ddot{\sigma}} &=& \mathcal{C}_\surf  36\pi n_\cl^2 r_\cl^2 \frac{\partial^2 \sigma_\surf }{\partial n_\cl^2} \, .
\end{eqnarray}

Note that the first derivative term, $\bar{K}_{\surf}^{\dd,\Dot{\sigma}}$, is expected to be small since $\partial \sigma_\surf / \partial n_\cl \propto x_\cl \approx 0$. In order to compare the above finite size contributions for Eq.~\eqref{eq:kanew} with Eq. (6.3) of Blaizot \cite{Blaizot80}, we shown in App.~\ref{sec:blaizotlike} how to write $\bar{K}_{A,\FS}$ in Blaizot notation.

\subsection{Sensitivity analysis}
\label{sec:sensitivity}

\begin{figure*}[t]
\centering
\includegraphics[scale=0.45]{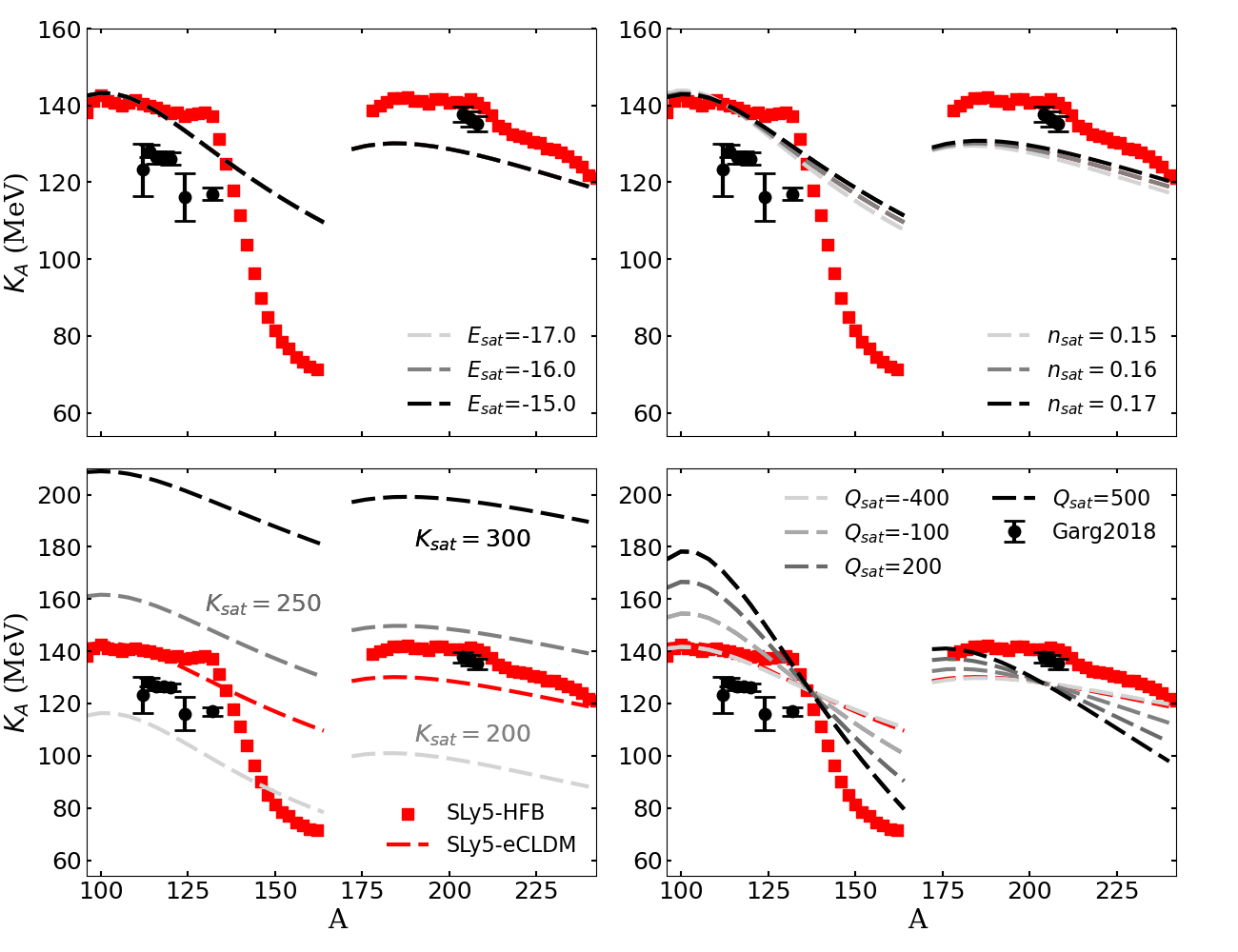}
\caption{Incompressibility for Sn and Pb isotopes. Black dots with error bars show the results of experimental data of Garg {\it et al.} \cite{GARG2018}. Red dashed lines (squares) shows the predictions from eCLDM (CHFB) with SLy5 interaction. Different line colors (light grey to black) show variation on isoscalar empirical parameters $E_\sat$ (top left), $n_\sat$ (top right), $K_\sat$ (bottom left) and $Q_\sat$ (bottom right). }
\label{fig:satEP}
\end{figure*}

\begin{figure}[t]
\centering
\includegraphics[scale=0.35]{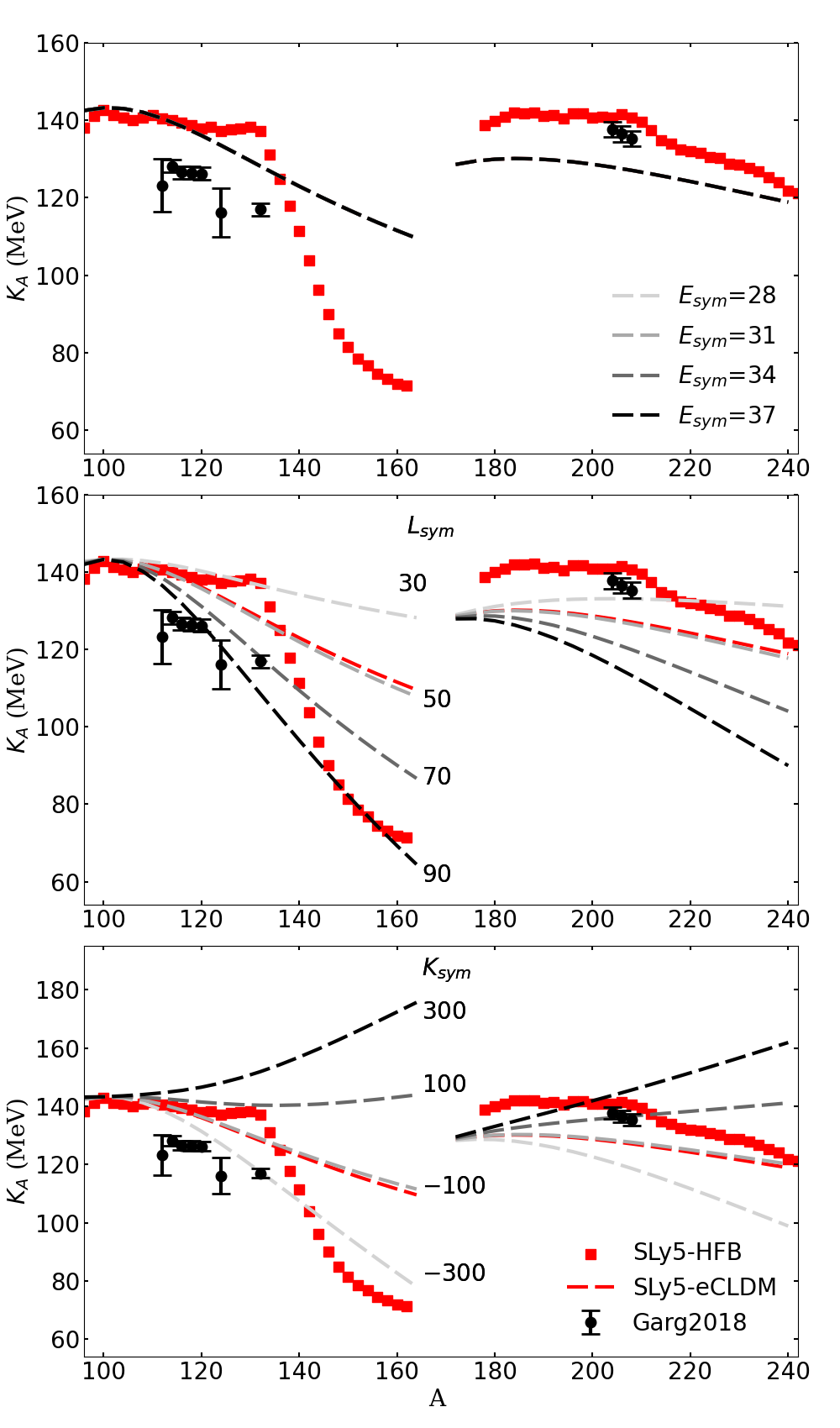}
\caption{Same as Fig. \ref{fig:satEP} but for variation on the isovector empirical parameters $E_\sym$ (top), $L_\sym$ (center) and $K_\sym$ (bottom). }
\label{fig:symEP}
\end{figure}

We analyse the impact of both the isoscalar NEPs ($E_\sat$, $n_\sat$, $K_\sat$ and $Q_\sat$) in Fig.~\ref{fig:satEP}, and the isovector NEPs ($E_\sym$, $L_\sym$ and $K_\sym$), in Fig.~\ref{fig:symEP}. The results obtained from the microscopic CHFB calculation based on the Skyrme SLy5 Hamiltonian~\cite{Chabanat98} are shown in red square for the two figures. The experimental data of Tab.~\ref{tab:KA}, are shown in black with their error-bars.

In Fig.~\ref{fig:satEP} we show the impact of the isoscalar parameters $E_\sat$ and $n_\sat$ (top), $K_\sat$ and $Q_\sat$ (bottom). The effects of $E_\sat$ and $n_\sat$ are very small and almost unnoticeable. However, the incompressibility modulus $K_\sat$ largely impacts $K_A$ with a positive correlation: the larger $K_\sat$ the larger $K_A$. The impact of $Q_\sat$ is also large but less linear: there is a crossing value for $A$ for which the impact of $Q_\sat$ is negligible. On the left of this crossing $A$, $Q_\sat$ is correlated with $K_A$ and on the right of it, it is anti-correlated. The red dashed line represents the eCLDM results using SLy5 Skyrme force. The value for $K_\sat$ which predict $K_A$ above the experimental data in Sn, predict $K_A$ below them in Pb. It is then difficult to fix accurately $K_\sat$ to reproduce experimental data in both Sn and Pb isotopes. The difficulty to reproduce Sn and Pb isotopes within the same nuclear force is indeed well know in the literature \cite{COLO200415,GARG200736,PATEL2012447}. However, it is  possible to use the NEP $Q_\sat$ which impact $K_A$, in a different way compared to $K_\sat$, as previously commented. To reconcile eCLDM with nuclear data, a low value for $Q_\sat$ is preferred.

We now analyse the impact of the isovector NEPs.
In Fig. \ref{fig:symEP} we plot eCLDM predictions assuming the Skyrme SLy5 Hamiltonian (red dashed lines), and then as for the isoscalar NEPs, we vary the NEPs one after another. As expected, these parameters do not impact $K_{A}$ in symmetric nuclei, and have an impact which increases as the isospin asymmetries increase. The impact of $E_\sym$ (top panel) is however invisible at the scale of the figure, while $L_\sym$ (middle panel) and $K_\sym$ (bottom panel) have larger impacts: $L_\sym$ is anti-correlated with $K_A$, while $K_\sym$ is correlated with $K_A$. As the isospin asymmetry of the isotopes increases, the impact of these isovector NEPs gets larger and larger. The larger decrease of $K_A$ as function of $A$ in Sn isotopes, is obtained for large values of $L_\sym$, but a low value of $K_\sym$ could also simulate the same effect.

None of the variations around the SLy5 Skyrme force seems to be preferred by the data. It is then difficult, from this sensitivity analysis, to detect which parameter set best reproduces the experimental nuclear data: the role of the different NEP is complex and the values which suggest a better description of the data, seem far from the SLy5 ones. In order to search for the best parameter set, it is then necessary to have a more global approach, where all the NEPS could be varied together, which is what we present in Sec.~\ref{sec:mcmc}.

\section{Expression for $x_\cl$ in finite nuclei}
\label{sec:xClExpression}

We follow the approach of Blaizot \cite{Blaizot80} and rewrite $x_\cl$ as follows. From the definition of the compressibility $\chi(n)$,
\begin{equation}
\chi = \frac{1}{n}\left(\frac{d P}{d n}
\right)^{-1}\, , \;\;\; \hbox{ we have } \;\;\; \frac{d P}{d n}= \frac{1}{n \chi}.
\label{eq:compress}    
\end{equation}
In $N=Z$ nuclei, $P=P_{A,SM}$, and by integrating~\eqref{eq:compress} from saturation ($n_\sat$) to equilibrium ($n_\cl$),
\begin{equation}
P_{A,SM}(n_\cl) - P_{A,SM}(n_\sat) =\int^{n_\cl}_{n_\sat}  \frac{1}{n \chi}dn \, .
\end{equation}
By definition $P_{A,SM}(n_\sat)=0$, and for $x_\cl$ is close to $n_\sat$ we approximate $\chi(n) \approx \chi (n_0)$ with $n_0\in [n_\sat, n_\cl]$, leading to
\begin{equation}
P_{A,SM}(n_\cl) \approx
\frac{1}{\chi(n_0)}\log \frac{n_\cl}{n_\sat} \approx
\frac{1}{\chi(n_0)}\frac{n_\cl - n_\sat}{n_\sat} \, .
\end{equation}
Since $K=9/(n\chi)$, we have
\begin{equation}
\frac{1}{\chi(n_0)} = \frac{n_0K_{A,SM}(n_0)}{9}
\approx \frac{n_\cl K_\sat}{9} , ,
\end{equation}
since $K_{A,\sm}$ is an increasing function of the density.
Finally, we obtain
\begin{equation}
x_\cl = \frac{n_\cl - n_{\sat}}{ 3n_{\sat}} = \frac{3}{ n_\cl K_{\sat}}P_{A,SM}(n_\cl) .   
\label{eq:n0nsat}
\end{equation}
Eq.~\eqref{eq:n0nsat} could be interpreted as the following: there is an equivalence between the density shift $x_\cl$ which is different from zero for densities different from $n_\sat$, as an effect of an external pressure $P_{A,SM}$, shifting the equilibrium density to a slightly different one. In finite nuclei, this extra-pressure is originating from the FS and isospin asymmetry terms, since $P_{A}(n_\cl) = 0$. We therefore deduce $P_{A,SM}(n_\cl) = - P_{A,\FS}(n_\cl) - \delta^2 P_{A,\sym}(n_\cl)$, and we can rewrite Eq.~\eqref{eq:n0nsat} as 
\begin{equation}
x_\cl = - \frac{3}{ n_\cl K_{\sat}}\left(P_{A,\FS}(n_\cl) + \delta^2P_{A,\sym}(n_\cl) \right).   
\label{eq:xClinPress}
\end{equation}

\section{Contributions to the incompressibility modulus within the Blaizot notations}
\label{sec:blaizotlike}

In the original notations of Blaizot \cite{Blaizot80}, the NEP where not used, but instead the third derivative of the energy density $\epsilon$. Using the original notations, we obtain for the Coulomb contribution,
\begin{eqnarray}
\tilde{K}_{A,\coul} &=& \frac{3}{5}\frac{Z^2e^2}{A R_A}\left(1 -\frac{ 27n_\sat^2}{K_\sat}\frac{d^3 \epsilon}{d n^3}\right) \, ,
\end{eqnarray}
and for the surface contribution
\begin{eqnarray}
\tilde{K}_{\surf}^{\cldm} &=& 16 \pi r^2_{\cl} \sigma_\surf\left(
1 + \frac{27}{2} \frac{n_{\sat}^2}{ K_{\sat}}   
\frac{d^3 \epsilon}{d n^3}\Bigr|_{n_\sat}
\right) \, , \\
\tilde{K}_{\surf}^{\dd,\Dot{\sigma}} &=& - 12\pi n_\cl r_\cl^2 \frac{\partial \sigma_\surf }{\partial n_\cl}\left( 
1 + 27 \frac{n_\sat^2}{K_\sat}\frac{d^3 \epsilon}{d n^3}\Bigr|_{n_\sat} \right) \, , \\
\tilde{K}_{\surf}^{\dd,\ddot{\sigma}} &=& 36\pi n_\cl^2 r_\cl^2 \frac{\partial^2 \sigma_\surf }{\partial n_\cl^2} \, .
\end{eqnarray}

Where the relation between the NEP and the third derivative of the energy density can be obtained using,
\begin{equation}
27 n_\sat^2\frac{d^3 \epsilon}{d n^3}\Bigr|_{n_\sat}=
9 K_\sat + Q_\sat .
\label{eq:3derivEps}
\end{equation}

\section{relation between the derivatives in $n_\cl$ and the ones in $R_A$ in the eCLDM}
\label{app:nclrcl_deriv}

In this section, we provide the relations between derivative as function of $n_\cl$ and as function of $R_A$, considering the conservation of the mass number $A=\frac 4 3 \pi R_A^3 n_\cl$. These relations are employed in finite nuclei, since the FS terms have an explicit dependence on $R_A$ while the bulk terms depend on $n_\cl$.

We have the following relations for the first order derivatives:
\begin{equation}
\frac{\partial }{\partial n_\cl}\Bigr|_{A}=-\frac{R_A}{3n_\cl}\frac{\partial }{\partial R_A}\Bigr|_{A}, \;\;\; \mathrm{\&} \;\;\;   
\frac{\partial }{\partial R_A}\Bigr|_{A}=-\frac{3 n_\cl}{R_A} 
\frac{\partial }{\partial n_\cl}\Bigr|_{A} \, ,
\end{equation}
and for the second derivative:
\begin{eqnarray}
\frac{\partial^2 }{\partial R_A^2}\Bigr|_{A} =
12\frac{n_\cl}{R_A^2}\frac{\partial}{\partial n_\cl}\Bigr|_{A}+9\frac{n_\cl^2}{R_A^2}\frac{\partial^2}{\partial n_\cl^2}\Bigr|_{A}
\label{eq:secDerEnerRad}
\end{eqnarray}

\bibliography{ref}
\end{document}